# Large Thermal Motion in Halide Perovskites

T. A. Tyson[1,*], W. Gao[2], Y.-S. Chen[3], S. Ghose[4] and Y. Yan[5,*]

[1]Department of Physics, New Jersey Institute of Technology, Newark, NJ 07102
[2]Department of Chemistry, University of South Florida, Tampa, FL 33620
[3]ChemMatCARS, University of Chicago and Advanced Photon Source, Argonne National Laboratory, IL 60439
[4]National Synchrotron Light Source II, Brookhaven National Laboratory, Upton, NY 11973
[5]Department of Chemistry and Environmental Science, New Jersey Institute of Technology, Newark, NJ 07102

*Corresponding Authors: T. A Tyson, e-mail: tyson@njit.edu and
Y. Yan, e-mail: yong.yan@njit.edu

## Abstract

Solar cells based on hybrid perovskites have shown high efficiency while possessing simple processing methods. To gain a fundamental understanding of their properties on an atomic level, we investigate single crystals of $CH_3NH_3PbI_3$ with a narrow transition (~5 K) near 327 K. Temperature dependent structural measurements reveal a persistent tetragonal structure with smooth changes in the atomic displacement parameters (ADPs) on crossing T*. We show that the ADPs for I ions yield extended flat regions in the potential wells consistent with the measured large thermal expansion parameter. Molecular dynamics simulations reveal that this material exhibits significant high asymmetries in the Pb-I pair distribution functions. We also show that the intrinsically enhanced freedom of motion of the iodine atoms enables large deformations. This flexibility (softness) of the atomic structure results in highly localized atomic relaxation about defects and hence accounts for both the high carrier mobility as well as the structural instability.

# I. Introduction

Perovskites belong to a class of systems with chemical formula $ABX_3$, where B atoms (transition metal ions) sits at the center of a simple cube, the X atoms (oxygen or any other atom of the same column of the periodic table) are at the faces and the A atoms are at the cube corners [1,2]. They exhibit a broad range of properties including ferroelectricity and complex magnetic and electronic properties [1,3]. Typically the A site is occupied by a rare earth or alkali type ion, the B site is occupied by a transition metal ion, and the X site is occupied by an oxygen or some other atom of the same column of the periodic. These systems are quite stable structurally with respect to defects on the A and B sites with tolerable defect levels in the A site as high as 20 % [4]. The flexibility of their structure enables the existence of multiple closely lying ground states, yielding quite complex behavior [5].

In parallel with the traditional inorganic $ABX_3$ systems, recently hybrid organic/inorganic halide perovskites have recently attracted extensive attention. Perovskites with the A site replace by organic cations and the X site replaced by halides were explored as high-efficiency solar cells [6] with the advantages of having low processing temperatures and simple solution method synthesis [7]. Most recently, motivated by the work on $CH_3NH_3PbI_3$ [8], intensive photovoltaic studies have been conducted. Systematic experiments have raised the efficiency value to ~20%, and it is expected that combined with standard silicon-based technologies, efficiencies approaching 30% can be achieved [2, 9]. An important feature of this system is the high absorption cross section for photons in the optical region and the high carrier mobility with electron-hole diffusion lengths approaching 150 μm [2,10]. Understanding these properties to enable improvement of stability and enhanced efficiency requires a detailed knowledge of the structure on multiple length scales.

Early heat capacity measurements revealed the complex nature of the structural response to temperature in the $CH_3NH_3PbX_3$ system [11]. Strong peaks corresponding to a transition from a high-temperature cubic to an intermediate temperature tetragonal phase followed by a transition to the low-temperature orthorhombic phase were found. Based on theoretical models, it was argued that the orientation of the C-N bonds in the $CH_3NH_3^+$ (MA) ion is progressively ordered as the temperature is reduced. More recently, with the observation of high efficiency and optimal photovoltaic properties in $CH_3NH_3PbI_3$, a broad range of structural studies have also been conducted, but primarily over a long length scale (micron scale).

Specifically, powder neutron diffraction measurements on $CH_3NH_3PbI_3$ between 100 and 352 K were conducted [12]. Structural analysis reveals that the MA ions are fully ordered in the orthorhombic phase (with space group Pnma) with the $NH_3$ groups aligned in the face of the perovskite cell and with the $PbI_6$ octahedra slightly distorted. The tetragonal phase was found to be in the I4/mcm space group in which the MA ions adopt one of four orientations along the (100) direction and equivalent directions. In the cubic phase with Pm-3m space group, the MA ions were found to be orientationally disordered. As the temperature is increased, the bonding between the $NH_3$ groups in MA and iodine in the framework is progressively reduced. Structural studies of the general class of hybrid perovskites [13] were carried out on single crystals. The high-temperature phase was found to adopt the P4mm tetragonal space group (at 400 K), below 333 K the intermediate temperature tetragonal phase was found to adopt the I4cm space group. Earlier single crystal x-ray diffraction measurements indicated that the intermediate phases corresponded to the I4/mcm space group [14]. More recent single crystal neutron diffraction measurements (343 K) in the high-temperature phase indicate two possible unit cells which can fit all Bragg peaks, the P4mm structure and rhombohedral structure (R-3m or R3m) [15]. Single crystal x-ray diffraction measurements at 298 K and 350 K yielded tetragonal and cubic structures (I4/mcm and Pm-3m space groups), respectively [16]. Single crystal x-ray diffraction measurements at 293 K on micron size crystals reveal a space group of I4cm [17] and further pair distribution function measurements at 350 K suggest that a tetragonal instantaneous structure on short length scales (I4cm, 2 to 8 Å) but cubic on longer length scales (Pm3m, 12 to 50 Å ) when dynamically differently oriented domains are averaged [18]. Off-centering of the Pb ions from the high symmetry sites was also observed. A broad range of density functional computations has been carried out on this class of materials. The stability of the possible phases was explored [19] revealing a very small difference between the free energy of the Pm-3m and P4mm as possible high-temperature structures. Recent pseudo-binary phase processing diagrams for $MAPbI_3$ between 313 and 463 K reveal intermediate and high-temperature tetragonal structures of I4/mcm and P4mm symmetry [20]. It was observed that obtaining the high-temperature phase required heating over a long time period.

Understating the intrinsic properties of these materials requires a broad range of studies on high-quality single crystals. To gain a fundamental understanding of their properties on an atomic level, we have prepared single crystals and have conducted heat capacity measurements between 170 K and 380 K, high-resolution single-crystal synchrotron x-ray diffraction measurements at room temperature, x-ray pair distribution function measurements (PDF) between 170 K and 445 K and x-ray absorption fine structure (XAFS) local structural measurements at room temperature. These measurements are essential to illustrate the unique structural properties of responsible for the high efficiency of these materials.

Molecular dynamics (MD) simulation using *ab initio* density functional methods were utilized to assess the structure at 200, 300, 400 and 500 K. Also, MD simulation at 300 K in the presence of combined MA and I defects were conducted. The simulations focused on a temperature range relevant to the natural operating temperature of solar cell devices. Our experiments reveal a transition with a narrow width (~5 K) occurring near 327 K with very weak hysteresis (~continuous transition). The room temperature single-crystal synchrotron x-ray diffraction measurements reveal a space group of I4cm with anomalously large and asymmetric thermal parameters for I1 and I2 transverse to the Pb-I bonds. The MD simulations show large asymmetry in the Pb-I pair peaks at all temperatures (200 K to 400 K) indicating anharmonic motion of the I atoms. Simulations of 300 K systems with defects at the MA and I site reveal that the lattice distortion about the I defect occupies a very restricted volume of space due to the lattice flexibility. X-ray pair distribution function measurements (at a 60 Ǻ length scale) reveal that a tetragonal structure exists over the full range studied (170 to 445) and yield a very large volume thermal expansion parameter. Anomalies are found in the thermal parameters near the transition (covering a range of more than 40 K). Modeling of the thermal parameters reveals potentials with very broad minima for the I2 and to a lesser extent the I1 ion. The Pb-I pair distribution extracted from XAFS measurements also indicated a very high degree of anharmonicity. Collectively, the results indicate that the I atoms can move considerably within the lattice at very low cost in energy. This high lattice flexibility enables the system to support a high density of defects each with limited spatial extension- leading to high charge mobility. At the same time the high flexibility also makes the system less structurally stable than standard oxide perovskites. On a local level, the sample has a mixed phase tetragonal/cubic structure for a broad temperature range.

# II. Results and Discussion

Figure 1 shows the heat capacity measurements of $CH_3NH_3PbI_3$ on warming and cooling with the inset revealing the peak at 329.2 K (T*). The hysteresis was found to be very weak (inset) and near the limit of the experimental measurements ~(shift less than 0.2 K). Also the width of the region where the heat capacity varies strongly is ~ 5 K. In addition, below that transition down to ~170 K, the heat capacity exhibits a nearly linear behavior. The straight line (guide to the eye) between 175 K and 310 K shows the highly linear behavior below T*. Formally, $C_p = C_v + n\,V\,T\alpha_V^2/\beta$, where $\alpha_V$ is the volume thermal expansivity and $\beta$ is the isothermal compressibility. We note that the difference between $C_p$ and $C_v$ is small in solids and that in the high temperature limit $C_p \propto 1 + \alpha_V T$ [21]. In the case of systems with large thermal expansion, linear variation is expected at high temperature. Thus the linear behavior of the heat capacity is consistent with large $\alpha_V$ values and anharmonic behavior.

Room temperature single-crystal synchrotron x-ray diffraction data were collected for short and long times and merged (eliminating any saturated peaks) to obtain high signal to noise data. This enables the determination of the atomic structure at the Pb, I, C and N sites. The space group I4cm was found (see data presented in Table I and Tables S1 and S2 (Supplementary Document)). The isotropic atomic displacement parameters (ADPs) of the I1 and I2 ions are found to be four times larger than those of the Pb ion. In Fig. 2 and Fig. S1 we see that thermal ellipsoids are highly extended transverse to the bonding direction in the Pb-I-Pb bond chain. These results indicate large motion of the I1 and I2 ions transverse to the Pb-I-Pb chains (see Table S1 for the anisotropic displacement parameters). We note that the orientation of the MA ion can also be extracted from the data. In Fig. 2, the two symmetry equivalent (mirror plane) N ions (N and N*) are shown indicating two possible orientations of the C-N bond with respect to the unit cell. No constraints were required on the C-N distances. Utilizing the extracted atomic positions, we calculated the charge density in a plane containing the Pb-I2-Pb bonds (See Fig. S2). The results indicate small charge build up between the the Pb and I2 sites, consistent with very weak Pb-I bonding.

To understand the motion of the ions in the lattice, MD calculations were conducted on a 2x2x1 unit cell at fixed temperature for T= 200, 300, 400 and 500 K. The radial distribution functions for atomic pairs are given in Fig. 3(a) and information on the simulations is presented in the supplementary text (Figs. S3 to S5). The pair distribution of the nearest neighbor Pb-I distances is given in Fig. 3(b) for all temperatures with fits to the simulations at 400 K (dashed line) and at 300 K (solid line). For the entire temperature range 200 to 400 K, the distribution function is asymmetric. The asymmetry in the radial distribution functions does not significantly change over this temperature range. The presence of this significant asymmetry indicates that the motion of the atoms is highly anharmonic as suggested by the heat capacity and the large ADPs in the single crystal measurements. The calculations also show that that hybrid perovskites are quite soft compared to traditional perovskite oxides. The $PbI_3$ cage has very low phonon modes revealed by the phonon density of states, computed from the velocity autocorrelation function [22] based on the molecular dynamics simulations, with peaks near 40 $cm^{-1}$ and 130 $cm^{-1}$ (due to Pb and I motion, respectively (see Fig. S4)). The molecular dynamics simulation also reveal a compression of the C-N bond in the MA ion when hosted in the perovskite lattice compared to isolated MA ions (Fig. S5).

To explore the dynamic behavior of the lattice on a nanoscale, single crystals of the sample were crushed to form a micron scale powder for large angle hard x-ray synchrotron powder diffraction with high momentum transfer. Data were collected between 170 K and 445 K, covering the region of the transition at T*. The system was modeled using the space group I4cm to search for tetragonal structures. A representative fit at room temperature is shown in Fig. S6(a) (Supplementary text) and the quality of the fit vs temperature is given in Fig. S6(b) as the $R_w$ parameter. Stability of the C-N bond provides a more robust indicator of the quality of the fits (Fig. S6(c)). All fits were conducted over the range 2.75 to 60 Å (nanoscale modeling in real space).

In Fig. 4(a) we show the ratio c*/a* ($\frac{c/2}{a\sqrt{2}}$) as a function of temperature. The figure gives the original a and c lattice parameters (lower left) and the volume versus temperature (upper right) as insets. The lattice parameters indicate that the system is tetragonal for the complete temperature range shown. Data points (red squares) from the work of Stoumpos *et al* [13] also fall close to the data on this work. From the volume vs temperature curve we obtain a value of $\alpha_V = 1.097(5) \times 10^{-4}\ K^{-1}$ which is extremely large and compared to very high values found for some crystalline organic materials [23].

From the extracted fit parameters we plot the ADPs vs temperature in Fig. 4(b), representing the I1 and I2 atoms. (Note that anisotropic parameters $U_{11}$ and $U_{33}$ were refined for all structural fits to data but the average $U_{iso}$ parameters are given in the figures and used in the potential model fits.) While the heat capacity changes rapidly over a ~5 K temperature range, we see that isotropic ADPs for I1 and I2 show smooth, broad variation over a range of ~40 K and ~80 K, respectively. The lack of an abrupt or discontinuous change in the ADP is consistent with a continuous phase transition. To understand the motion of the atoms, the extracted ADPS were fit to an effective potential $V(u) = \alpha u^2 + \gamma u^4$ (see Ref. [ 24 ]) with the isotropic ADP for any temperature given by $U(Å^2) = \int_0^\infty x^4\, Exp\left[-\frac{V(x)}{k_B T}\right] dx \,/ \int_0^\infty x^2\, Exp\left[-\frac{V(x)}{k_B T}\right] dx$ assuming that the motion of the atoms follows a Boltzman distribution. The parameters $\alpha$ and $\gamma$ are, respectively, the force constant and quartic anharmonic contribution for motion of the atom with respect to its equilibrium position. Data between 180 and 280 K (below T*) for the Pb, I1 and I2 ADPS were fit to determine the potential function parameters (written as $\alpha_i$ and $\gamma_i$). The fits are displayed in Figs. S7 and S8. In Fig. 4(c) we plot the potentials for Pb, I1 and I2. These functions give information about the physical motion of the atoms. I2 displacements of the ordered of 0.2 Å can be excited with energy $k_B T$ near room temperature. Large values of the $\gamma_i$ parameters reveal the high anharmonic nature of the motion. The I2 atoms and to a lesser extent the I1 atoms move in broad shallow wells dominated by motion transverse to the bonds with respect to their near neighbor Pb sites.

The behavior of Pb-I bonding pair was determined by room temperature XAFS measurements at the Pb L3 edge. The room temperature data exhibits just one clear peak (Fig. S9) in the Fourier transform of the extracted fine structure signal, which corresponds to the distribution of Pb-I bonds. Higher order peaks such as those for Pb-Pb are not evident due to very weak correlation of the motion of these atom pairs- confirming the soft nature of the lattice. Fits of the data with a distribution based on a split Gaussian peak reveal high asymmetry in this bond. Additional modeling with an anharmonic pair potential $V(r) = \frac{\alpha}{2}(r - r_0)^2 + \frac{\beta}{6}(r - r_0)^3 + \frac{\gamma}{24}(r - r_0)^4$ [25] enables the determination of the force constant and anharmonic parameters for this bond. The resulting bond distributions from both methods as well as the extracted potential function are given in Fig. 5.

These results can be integrated to present a coherent picture of the behavior of the system over the temperature range 170 to 445 K, when combined with previously published results. It is well established that progressive ordering of the MA ions occurs on cooling from high temperature through both transitions in $CH_3NH_3PbI_3$ [26]. We found no discontinuous changes, as seen in the lattice parameter (volume) variation with temperature (Fig. 4). Hence, combined with the weak hysteresis observed in the heat capacity measurements presented here we are led to assert that the transition at T* is a continuous transition. An important observation from the PDF measurements is that while the width of the peak in heat capacity is ~5 K , changes in the x-ray diffraction ADPs occurs over regions up to ~60 K wide. We note also that the resistivity was found to change over a broad temperate range [13] without a discontinuity. It implies that on the timescale and distance scale (60 Å) of the PDF measurements the sample exists as a mixture of the high temperature and low-temperature phases significantly below T* and significantly above T* with the tetragonal phase observable up to the highest temperature measured, ~445 K. Hence the structure which hosts the high-efficiency photovoltaic properties is a pure tetragonal or mixed tetragonal and cubic phase. We note that since the transition at T* is second order (more accurately described as continuous), that the symmetry of the high-temperature phase must obey a group-subgroup relation relative to the low-temperature phase.

Over the temperature range studied (170 to 445 K), one can consider each unit cell (I4cm space group) as having a pseudo spin defined by the unique symmetry axis (c-axis) and which yields a net polarization direction in the cell. At high temperatures, the MA ions are randomly oriented while for low temperatures the restricted motion breaks the cell symmetry. The system can then be considered as a three-dimensional Ising lattice [27]. Such an Ising system has a characteristic asymmetric heat capacity which peaks at the phase transition temperature strongly when dominated by nearest neighbor interactions [28]. The correlation between unit cells with given "spin" directions give the observed variation of heat capacity in the vicinity of T*. This transition, similar to a magnetic, the transition is continuous. At high temperature, this model supports transitions in a phase characterized by the formation of the low-temperature phase regions embedded in the high-temperature phase above away T*. While below T* it supports the high-temperature phase embedded in the low-temperature phase. This is consistent with the combined heat capacity and ADPs measured by PDF.

To under understand the election transport in this system we note that as in the case of the classic manganite oxide perovskites [29], the Pb and I sites provide the network along which transport occurs. The hopping of electrons from nearest neighbor Pb to Pb sites via the I sites is related to the overlap of the Pb 6s and I 6p wave functions. The Pb-I1-Pb bond is along the c-axis (180º), but Pb-I2-Pb bond angle varies with temperature. Following the methods of Harrison [30], we can express the hopping integral as $t \sim |V_{sp\sigma}|^2 \cos^2(\theta)$ (Fig. 6(a)). As seen from the relation between the extracted bond angle and temperature from the PDF data, the overlap integral increases with temperature and hence tracks the observed resistivity reduction (conductivity increase) [12] as temperature is increased. The MD simulations of the Pb-I-Pb bond distance with time also shows an increase in average value with temperature but with large fluctuations (Fig. 6(b)) at high temperatures. We note that in the presence of defects at both the I and MA sites the behavior of the Pb-I-Pb bond distribution is qualitatively the same (Fig. 6(b)).

The connection of the large possible displacement of the I ions at low energy cost can be related to the high observed charge mobility. The large motion of these ions makes possible charge separation. This separation has been shown to be responsible for a large dielectric constant observed over a broad frequency range [31]. Theoretical predictions on these systems point to the Pb-I stretching and Pb-I-Pb rocking modes dominating the atomic contributions to the dielectric constant for frequencies below 111 $cm^{-1}$ [32]. The large dielectric constant in these materials is made possible by the I motion provides a high degree of screening for charged defects. The screening of the defects as in the case of SiC [33], reduces impurity scattering significantly possibly leading to the very large carrier mobility. It is noted that the mobility is not as high as semiconductors such as GaAs. Hence these systems incorporate, a balance between the property of high dielectric constant for defect screening and increased electron-phonon scattering.

We illustrate the deformation of the lattice about defects tin Fig. 6(c). For the region around an I defect, a section of the unit cell is shown for the average structure (from MD simulations at 300 K) with one MA and one I defect present. Atoms relax away from missing I atoms. The large dark arrow indicates the position of the missing I atom and the green arrows give the displacement of the atoms relative to the average structure without defects. The longest displacement arrows in the figure correspond to 0.79 Å. The softness of the lattice which underlies the high anhramonicity enables deformation of the lattice in response to the defect but limits its extension to a very small region of space (shown by the loop in Fig. 6(c)). The deformation about the I defect can be roughly represented by a sphere of ~~ 10 Å diameter.

## III. Summary

Experiments on single crystal from the same batch have been carried out using multiple structure related methods. Measurements covering the temperature range ~ 170 K to ~445 K were compared with molecular dynamics calculations. Heat capacity measurements reveal a continuous transition near ~330 K (T*)- which is reasonably close to solar cell operating temperatures. Structural studies near the operating temperature region were conducted. Local structural PDF measurements reveal a tetragonal structure over the full range with smooth broad smooth changes in the atomic displacement parameters (ADPs) at T* extending significantly above and below T*. Fits to ADPs make evident large flat regions in the potential wells in which the I1 and I2 are moving. High anharmonic behavior is found with an extremely large thermal expansion parameter $\alpha_V$. The reduction in the resistivity at high temperature is shown to correlate with the temperature variation in the Pb-I2-Pb bond angles. The softness of the lattice which underlies the high anhramonicity, enable deformation of the lattice in response to the defect but limits its extension it to a very small region of space yielding a material with resilient high carrier mobility in the presence of defects. and requires theoretical models which incorporate large anharmonic motion of the atoms.

The soft potential wells, in which the I and Pb ions sit, enhance their electrical properties. However, it may also make the materials structurally unstable and mechanically soft. Supporting the arguments presented here, recent theoretical work explored candidate compounds in this class based on Cs at the A site show that the iodide type perovskite is intrinsically unstable with quite low decomposition enthalpy [34]. Overall, the results indicate that this class of hybrid perovskite has distinctly different physics from oxide perovskites and requires theoretical models which incorporate large anharmonic motion of the atoms. Our structural explorations at temperatures near perovskite solar cell operation temperature demonstrated will pave the way to understand the impact of large thermal motion on the critical properties, i.e. high carrier mobility, of $MAPbI_3$ perovskite materials. Our work in this direction will significantly stimulate wider exploration of fundamental structural correlations with the high performance of hybrid perovskite solar cell materials.

## IV. Experimental and Modeling Methods

Details of the synthesis and experimental and modeling methods are given in the supplementary document. High-quality single crystals of $CH_3NH_3PbI_3$ were prepared in gamma-butyrolactone. Specific heat measurements were conducted on warming from 300 K to 360 K and on cooling from 360 K to 170. All experiments were conducted on samples from the same preparation batch. Approximately 10 minutes was required to measure each temperature data point. Hence the system was never in a quenched state. Synchrotron single crystal x-ray diffraction measurements were conducted on ~15 $\mu$m diameter crystals. Pair distribution function experiments (conducted at similar collection time per temperature point as the specific heat measurements) and XAFS measurements were conducted on crushed single crystals (500 mesh powders). MD simulations were conducted utilizing the VASP code implementing projector-augmented wave (PAW) potentials. The datasets generated during and/or analysed during the current study are available from the corresponding author on reasonable request.

## V. Acknowledgments

This work is supported by DOE Grant DE-FG02−07ER46402. Y. Y. is supported by NJIT faculty startup funds. This research used resources of the National Synchrotron Light Source, a U.S. Department of Energy (DOE) Office of Science User Facility operated for the DOE Office of Science by Brookhaven

National Laboratory under Contract No. DE-AC02-98CH10886. Single crystal x-ray diffraction measurements were performed at ChemMatCARS Sector 15, which is principally supported by the National Science Foundation/Department of Energy under Grant NSF/CHE-1346572. Use of the Advanced Photon Source was supported by the U.S. Department of Energy, Office of Science, Office of Basic Energy Sciences, under Contract No. DE-AC02-06CH11357. The Physical Properties Measurements System was acquired under NSF MRI Grant DMR-0923032 (ARRA award). This research used resources of the National Energy Research Scientific Computing Center, a DOE Office of Science User Facility supported by the Office of Science of the U.S. Department of Energy under Contract No. DE-AC02-05CH11231. We are indebted to K. H. Ahn and C. Dias of the NJIT physics department for help theoretical discussions.

## Author Contributions

T. A. T. and Y. Y. prepared the manuscript. Sample synthesis was conducted by Y. Y. Single crystal diffraction measurements were performed by W. G. and Y.-S. Chen and PDF data collection was done by S. Ghose. Heat capacity measurements were conducted by T. A. T. Single crystal diffraction, PDF, and XAFS data reduction and analysis were conducted by TAT in addition to the DFT and MD simulations.

## Competing Financial Interests

The authors declare no competing financial interests.

**Table I. Structural Parameters from Single Crystal Refinement at Room Temperature**

| **Atoms** | **x** | **y** | **z** | Ueq ($Å^2$)×$10^3$ |
|---|---|---|---|---|
| Pb | 0 | 0 | 0.22462(2) | 39.8(8) |
| I1 | 0 | 0 | 0.4718(13) | 121 (3) |
| I2 | 0.2780(6) | 0.2220(6) | 0.2207(17) | 112 (3) |
| N | 0.4168(42) | 0.0833(42) | 0.4520(35) | 26 (11) |
| C | 1/2 | 0 | 0.3677(78) | 76 (24) |

N site occupancy = 50%

$U_{ij}$ (Pb) 0.0390(9) 0.0413(10) 0.0000 0.0000 *

$U_{ij}$ I1) 0.1666(44) 0.0298(25) 0.0000 0.0000

$U_{ij}$ (I2) 0.0929(25) 0.1506(64) -0.0548(28) -0.0381(60)

a = 8.9406 (13) Å, c = 12.6546 (25) Å

BASF twin parameter: 0.47(12)
(Racemic twinning parameter indicating ~equal up and down polarization domains)

* Atomic displacement parameters $U_{ij}$ ($Å^2$) are in the order: $U_{11}$, $U_{33}$, $U_{12}$, and $U_{23}$. Space group I4cm. See supplementary document for full table with experimental details (Table S1).

**Fig. 1.** Heat capacity measurement of $CH_3NH_3PbI_3$ between 175 and 380 K on warming and cooling. The inset shows that only very weak hysteresis occurs at transition T*. A straight line (guide to eye) between 175 K and 310 K shows the highly linear behavior below the T*

**Fig. 2.** Crystal structure of $CH_3NH_3PbI_3$ from a room temperature structural solution with I4cm space group. Note the large thermal ellipsoids on the I1 and I2 sites. It is found that these ions exhibit high anharmonic behavior. Note that N and N* are 50% occupied sites for nitrogen in the methylamine ion. (See also supplementary data document Fig. S1 and Table S1)

**Fig. 3.** (a) MD-derived pair correlation functions at 300 K for the Pb-I, Pb-H, I-I, Pb-C, Pb-N and Pb-Pb pairs. Note the broad width of the Pb-I peak and I-I peak in the host lattice. (b) MD-derived radial distribution functions for Pb-I peak at 200, 300, 400 and 500 K. Fits to the MD simulation with a left ($\sigma_{left}$) and right ($\sigma_{right}$) sided Gaussian (continuous at the peak) were used to assess the asymmetry. Large asymmetry persist for the entire temperature range with $Asym = \frac{|\sigma_{left}-\sigma_{right}|}{\sigma_{left}+\sigma_{right}}$ values of 0.278(5) and 0.226(3) at 300 K and 500 K, respectively. This asymmetry indicated a high degree of anharmonicity in the system at the I sites.

**Fig. 4.** Results from pair distribution fitting over the range 180 K to 450 K based on the I4cm model. The system was found to be tetragonal for the entire temperature range (See supplementary document Fig. S6). Panel (a) gives a plot of the ratio of c*/a* indicating that the system is tetragonal for the entire temperature range. The red squares are data taken from Stoumpos *et al* [13]. The lower left inset gives the lattice parameters, and the upper inset gives the volume vs temperature yielding a very large volume expansion coefficient $\alpha_V = 1.097(5) \times 10^{-4}\ K^{-1}$. The extracted atomic displacement factors for the I sites

(I1 and I2, see Fig. 2) are shown in (b). Note that there is a smooth transition with temperature. In (c) the extracted single particle potentials representing the motion of the I atoms about the equilibrium positions are shown. The large value of the fourth order coefficient $\gamma$ indicates very strong anharmonic behavior (See also supplementary data document Fig. S7). Not that the enhanced anharmonic behavior of the I2 site relative to the I1 site.

**Fig. 5.** (a) Radial pair distribution for Pb-I site obtain from fits to Pb L3 edge XAFS data (see supplementary data Fig. S9). A fit using an asymmetric Gaussian function was utilized. The asymmetry in this fit as well for a classical potential mode (dashed line) reveals that the pair correlation of Pb-I is anharmonic. (b) Classical V(r) for generating the bond distribution which best fits the XAFS data. Note the non-negligible coefficients for the high order terms (third and fourth order).

**Fig. 6.** (a) Pb-I2-Pb bond angles extracted from PDF fits as a function of temperature between 170 and 445 K. The inset shows the bond angle dependence of net overlap integral for hopping across the full bond. (b) Temperature dependence of the Pb-I_Pb bond angle as a function of time showing the large thermal fluctuation for the 500 K simulation (green line) compared to the 300 K simulation (black line). The thin line is for the 300 K model with and one MA and one I defect in the same cell (shown in panel (c)). (c) Section of unit cell for averaged structure (from MD simulations at 300 K) with a MA and a I defect showing region around I defect. Atoms relax away from missing I atoms. The large dark arrow indicates the position of the missing I atom and the green arrows give displacements of the atoms relative to the average structure without defects. The longest displacement arrows correspond to 0.79 Å.

**Fig. 1.** Tyson *et al.*

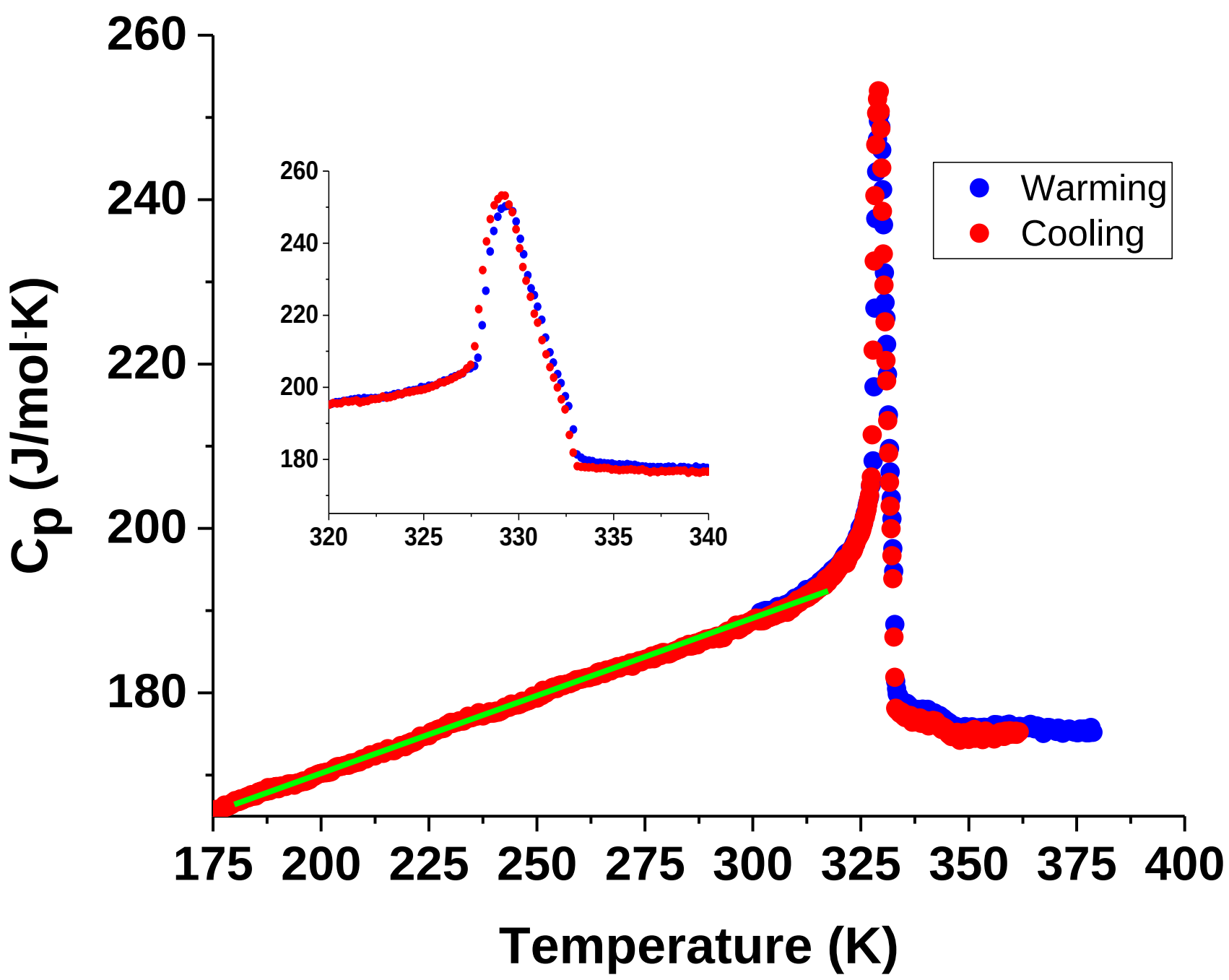

**Fig. 2.** Tyson *et al.*

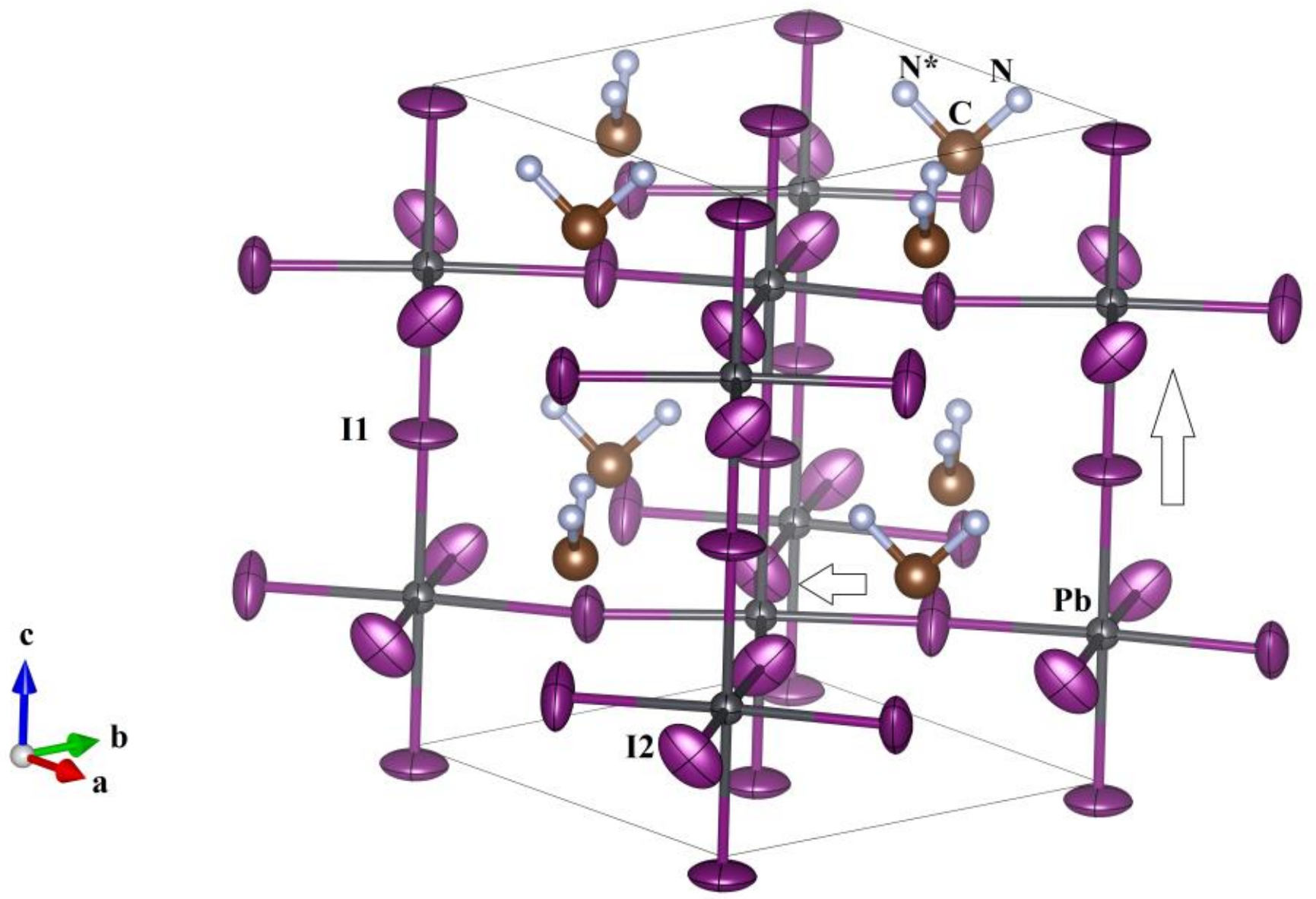

**Fig. 3.** Tyson *et al.*

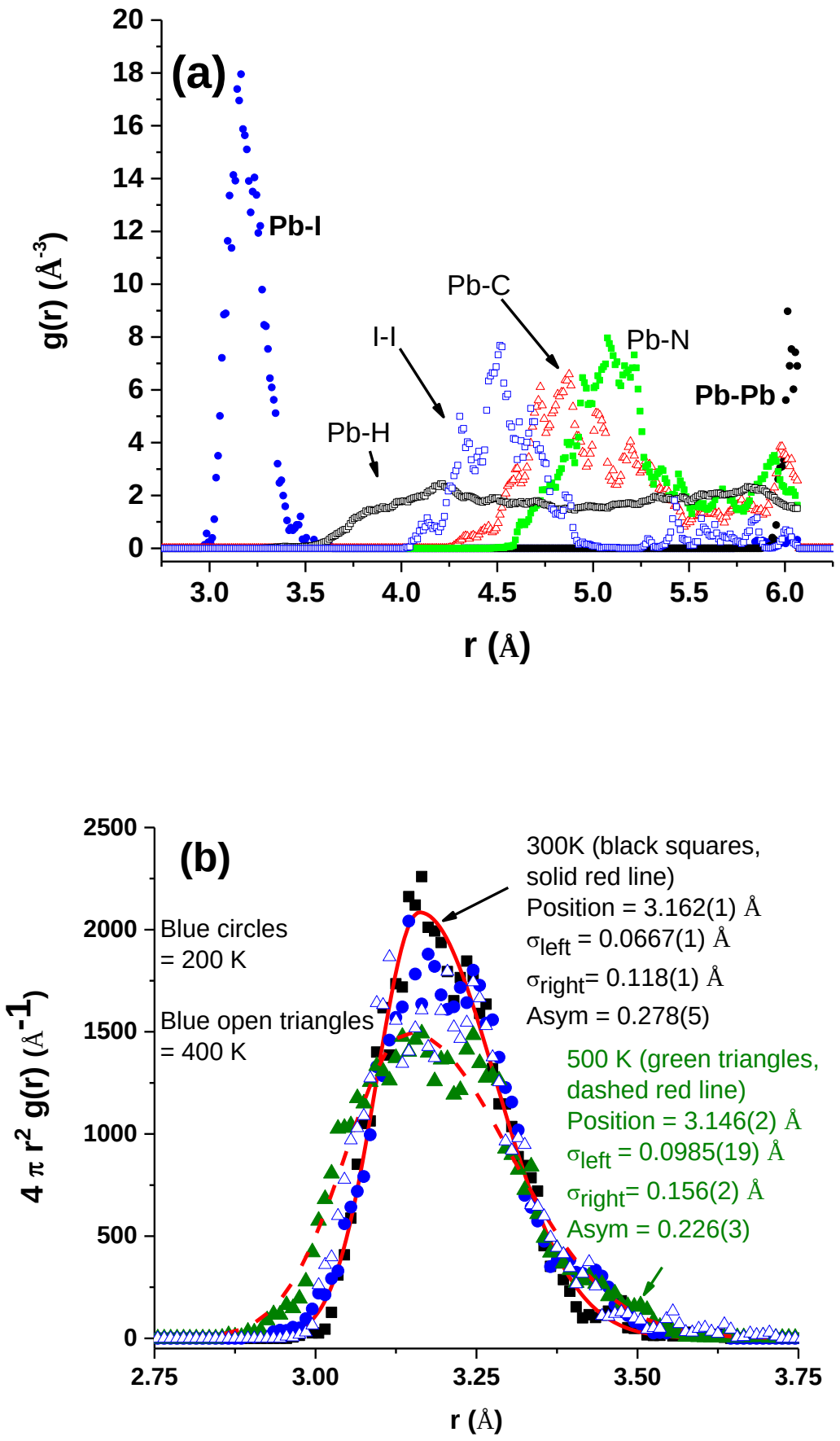

**Fig. 4.** Tyson *et al.*

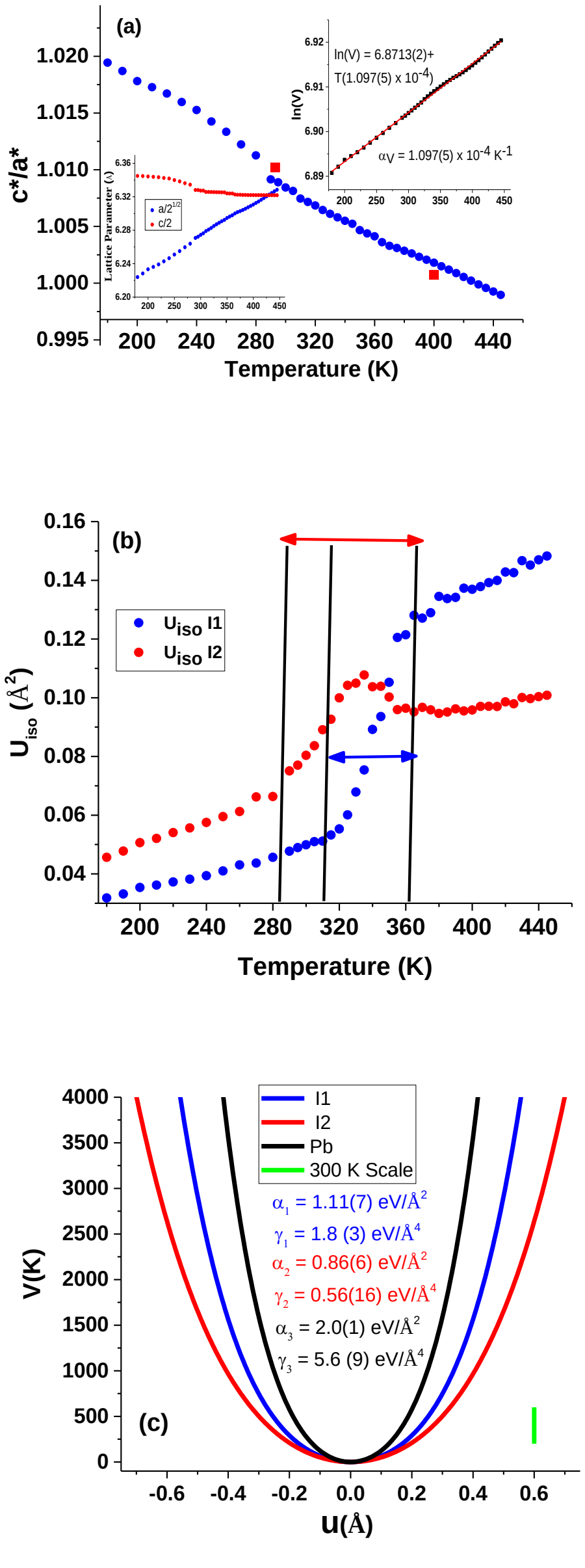

**Fig. 5.** Tyson *et al.*

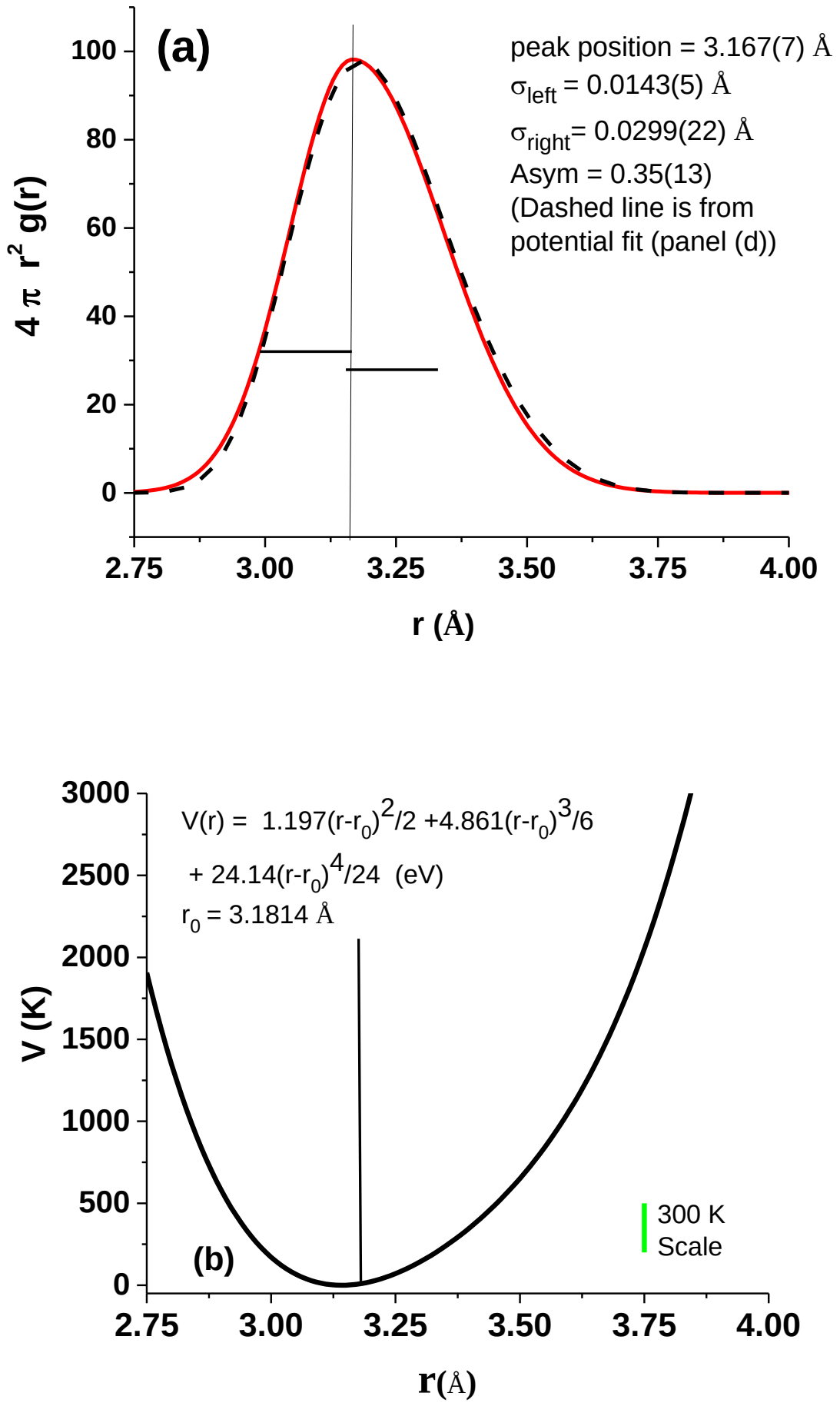

**Fig. 6.** Tyson *et al.*

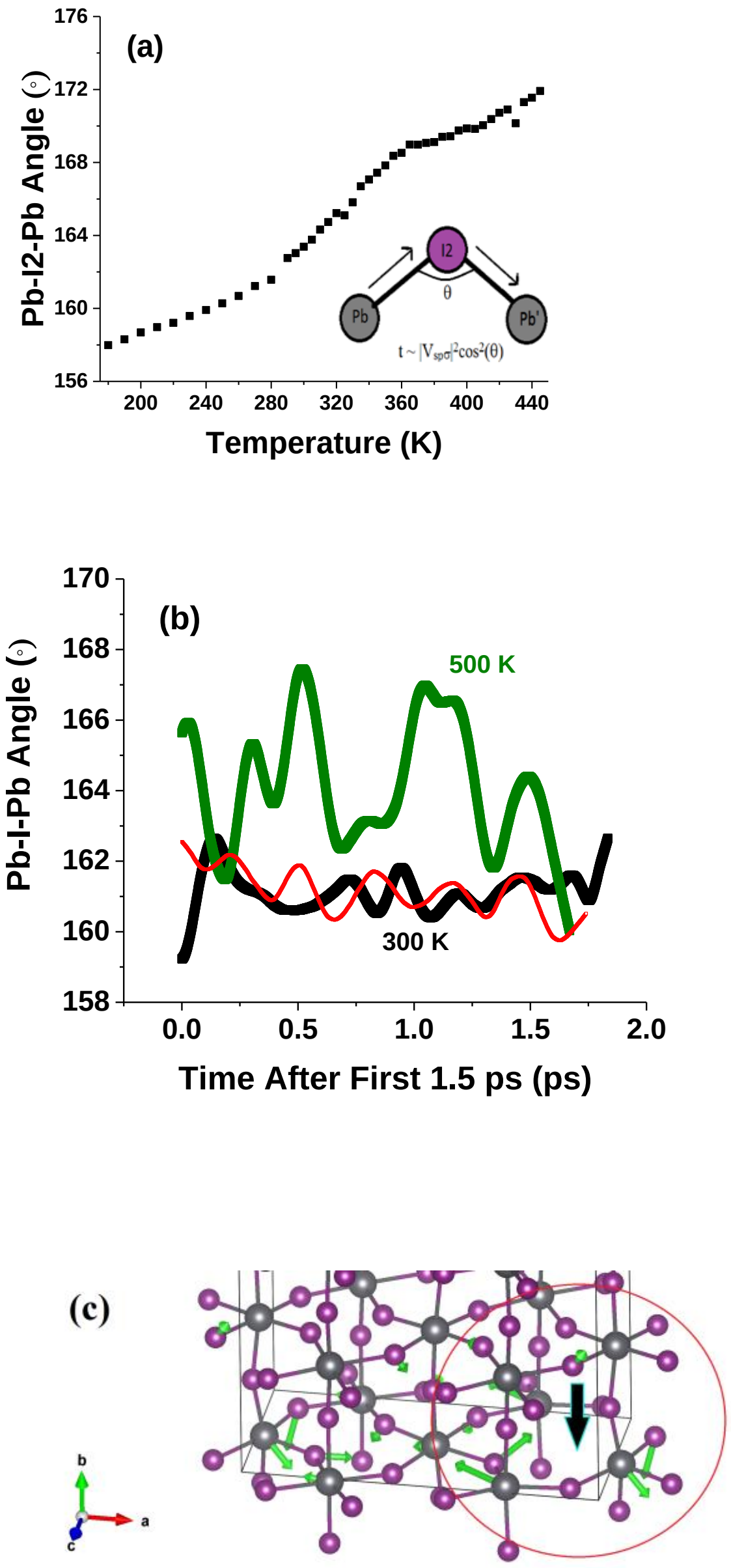

# References

[1] Moure, C. & O. Peña, O. Recent advances in perovskites: Processing and properties. Prog. *Solid State Chem*. **43**, 123-148 (2015) and references therein.

[2] Chilvery, A. K. *et al*. Perovskites: transforming photovoltaics, a mini-review. *J. Photonics Energy* **5**, 057402-1-057402-14 (2015) and references therein.

[3]Scott, J. F. Room-temperature multiferroic magnetoelectrics. *NPG Asia Mater*. **5**, 1-11 (2013).

[4] Bosak, A., Dubourdieu, C., Audier, M., Sénateur, J. P. & J.Pierre J., Compositional effects on the structure and magnetotransport properties of lacunar $La_{1-x}$ $MnO_{3-d}$ films (x> 0) grown by MOCVD. *Appl. Phys. A* **79**, 1979-1984 (2004).

[5] Ahn, K. H., Seman, T. F., Lookman, T. & Bishop, A. R., Role of complex energy landscapes and strains in multiscale inhomogeneities in perovskite manganite. *Phys. Rev. B Condens. Matter Mater. Phys*. **88** (14) 144415-1-144415-15 (2013).

[6] Saparov, B. & Mitzi, D. B., Organic–inorganic perovskites: structural versatility for functional materials design. *Chem. Rev.* **116** (7), 4558-4596 (2016) and references therein.

[7] Mitzi, D. B., Chondroudis, K., & Kagan, C. R. Organic-Inorganic Electronics. *IBM J. Res. & Dev.* 45, 29-45(1999) and reference therein.

[8] Lee, M. M. Teuscher, J., Miyasaka, T. , Murakami, T. N. & H. J. Snaith Lee, M. M. Teuscher, J., Miyasaka, T. , Murakami, T. N. & Snaith, Efficient hybrid solar cells based on meso-superstructured organometal halide perovskites. H. J. *Science* **338**, 643-647 (2012).

[9] Ye, M., Hong, X., Zhang, F. & Liu, X. Recent advancements in perovskite solar cells: flexibility, stability and large scale. *J. Mater. Chem.* A **4** (18), 6755-6771 (2016) and references therein.

[10] Dong, Q. *et al.* Solar cells. Electron-hole diffusion lengths > 175 μm in solution-grown $CH_3NH_3PbI_3$ single crystals. *Science* **347**, 967-970 (2015).

[11] Onoda-Yamamuro, N. , Matsuo, T. & Suga, H. Calorimetric and IR spectroscopic studies of phase transitions in methylammonium trihalogenoplumbates (II) *J. Chem. Phys. Solids* **51**, 1383-1395 (1990).

[12] Weller, M. T., Weber, O. J., Henry, O. J., Di Pumpo, A. M. & Hansen, T. C. Complete structure and cation orientation in the perovskite photovoltaic methylammonium lead iodide between 100 and 352 K. *Chem. Commun.* **51** (20), 4180-4183 (2015).

[13] Stoumpos, C. C., Malliakas, C. D., & Kanatzidis, M. G., Semiconducting Tin and Lead Iodide Perovskites with Organic Cations: Phase Transitions, High Mobilities, and Near-Infrared Photoluminescent Properties. *Inorganic. Chem.* **52**, 9019-9038 (2013).

[14] Kawamura, Y., Mashiyama, H. & Hasebe, K. Structural Study on Cubic–Tetragonal Transition of CH3NH3PbI3*J. Phys. Soc. Jpn*, **71**, 1694-1697 (2002).

[15] Baikie, T. *et al*. A combined single crystal neutron/X-ray diffraction and solid-state nuclear magnetic resonance study of the hybrid perovskites $CH_3NH_3PbX_3$ (X = I, Br and Cl). *J. Mater. Chem* A **3**, 9298-9307 (2015).

[16] Ren, Y., Oswald, I. W. H., Wang, X., McCandless, G. T. and Chan, J. Y. Orientation of Organic Cations in Hybrid Inorganic–Organic Perovskite $CH_3NH_3PbI_3$ from Subatomic Resolution Single Crystal Neutron Diffraction Structural Studies. *Cryst. Growth & Design*. **16**, 2945-2951 (2016).

[17] Dang, Y. *et al.* Bulk crystal growth of hybrid perovskite material $CH_3NH_3PbI_3$. *Cryst. Eng. Chem. Comm*. **17**, 665-670 (2015).

[18] A. N. Beecher *et al.* Direct Observation of Dynamic Symmetry Breaking above Room Temperature in Methylammonium Lead Iodide Perovskite. *Acs Energy Let*t. **1**, 880-887 (2016).

[19] Ong, K. P. , Goh, T. W., Xu, Q. & Huan, A. Structural Evolution in Methylammonium Lead Iodide $CH_3NH_3PbI_3$. *J. Phys. Chem. A* **119**, 11033-11038 (2015).

[20] Song, Z.,Watthage, S. C., Phillips, A. B., Tomkins, B. L., Eiilingston, R. J. & M. J. Heben, Impact of Processing Temperature and Composition on the Formation of Methylammonium Lead Iodide Perovskites. *Chem. Mater.* **27**, 4612-4619 (2015).

[21] Anderitsos, E. I. *et al.* The heat capacity of matter beyond the Dulong–Petit value. *J. Phys. Cond. Mat.* **25**, 235401_1-235401_6 (2013).

[22] Lan, T., Li, C. W. , Niedziela, J. L., Smith, H., Abernathy, D. L., Rossman, G. R., & Fultz, B. Anharmonic lattice dynamics of $Ag_2O$ studied by inelastic neutron scattering and first-principles molecular dynamics simulations. *Phys. Rev. B* 89, 054306_1 - 054306_10 (2014).

[23] Das, D., Jacobs, T. & Barbour, L. J. Exceptionally large positive and negative anisotropic thermal expansion of an organic crystalline material. *Nature Materials* **9**, 37 (2010).

[24] Sakata, M., Harada, J., Cooper, M. J., & Rouse, K. D. A neutron diffraction study of anharmonic thermal vibrations in cubic CsPbX3. *Acta. Cryst.* **A36**, 7-15 (1980).

[25] Ignatov, A. Y. , Dieng, L. M., Tyson, T. A., He, T. & Cava, R. J. Observation of a low-symmetry crystal structure for superconducting $MgCNi_3$ by Ni K-edge x-ray absorption measurements. *Phys. Rev. B* **67**, 064509_1-064509_7 (2003).

[26] Frost, J. M. & Walsh, A. What Is Moving in Hybrid Halide Perovskite Solar Cells? *Acc. Chem Res.* **49**, 528-535 (2016) and references therein.

[27] Binney, J. J., Dowrick, N. J., Fisher, A. J. & Newman, M. E. J. , *The Theory of Critical Phenomena*, (Clarendon Press, Oxford, 1992).

[28] Kozlovskii, M. P. & Pylyuk, I. V. Entropy and specific heat of the 3D ising model as functions of temperature and microscopic parameters of the system. *Phys. Stat. Sol.* **183**, 243-249 (1994).

[29] Coey, J. M. D. & Viret, M. Mixed-valence manganites *Adv. Phys.* **48**, 167-293 (1999).

[30] Harrison, W. A. *Electronic Structure and Properties of Solids*, (Dover Publications, Ann Arbor, 1989).

[31] Hoque, M. N. F. *et al*. Polarization and Dielectric Study of Methylammonium Lead Iodide Thin Film to Reveal its Nonferroelectric Nature under Solar Cell Operating Conditions. *ACS Energy Letter*s **1**, 142-149 (2016).

[32] Perez-Osorio, M. A. *et al.* Vibrational Properties of the Organic–Inorganic Halide Perovskite $CH_3NH_3PbI_3$ from Theory and Experiment: Factor Group Analysis, First-Principles Calculations, and Low-Temperature Infrared Spectra. J. Phys. Chem. C **119**, 25703-25718 (2015).

[33] Rengel, R., Iblasias, J. M., Pascual, E. & Martin, M. J. Effect of charged impurity scattering on the electron diffusivity and mobility in graphene. *J. phys. Conf. Ser*. **647**, 012046_1-012046_4 (2015).

[34] Zhao, X.-G. *et al.* Design of Lead-Free Inorganic Halide Perovskites for Solar Cells via Cation-Transmutation. *J. Am. Chem. Soc.* **139**, 2630-2638 (2017).

# Large Thermal Motion in Halide Perovskites (Supplementary Document)

T. A. Tyson[1,*], W. Gao[2], Y.-S. Chen[3], S. Ghose[4] and Y. Yan[5,*]

[1]Department of Physics, New Jersey Institute of Technology, Newark, NJ 07102
[2]Department of Chemistry, University of South Florida, Tampa, FL 33620
[3]ChemMatCARS, University of Chicago and Advanced Photon Source, Argonne National Laboratory, IL 60439
[4]National Synchrotron Light Source II, Brookhaven National Laboratory, Upton, NY 11973
[5]Department of Chemistry and Environmental Science, New Jersey Institute of Technology, Newark, NJ 07102

*Corresponding Authors: T. A Tyson, e-mail: tyson@njit.edu and
Y. Yan, e-mail: yong.yan@njit.edu

# I. Experimental and Modeling Methods

To synthesize single crystal of $CH_3NH_3PbI_3$, 3 mmol $PbI_2$ and 3 mmol $CH_3NH_3I$ were prepared in gamma-butyrolactone (3mL, targeting for 1mol/L solution). This mixture was heated up to 60 °C to completely dissolve any insoluble residue. The solutions were quickly filtered using a PTFE filter with 200 nm pore size. Two milliliters of the filtrate were placed in a vial which was then kept in an oil bath undisturbed at 110 °C for perovskites single crystal growth. All procedures were carried out under ambient conditions and a humidity of 55–57%. The crystals used for measurements were grown for 5 hrs. The obtained single crystals were dark with well- defined and highly reflective facets. For experiments involving powders, single crystals were ground into fine powders (below 500 mesh). All experiments were conducted on crystals from the same batch.

Specific heat measurements were conducted on warming from 300 K to 360 K and on cooling from 360 K to 170 K using the relaxation method in a Quantum Design PPMS system. Temperature steps of 0.2 K were utilized in the region of the transition near 329 K and each temperature point was measured three times and the average result is reported (Fig. 1). Approximately 10 minutes was required to measure each temperature data point. Hence the system was never in a quenched state.

Synchrotron single crystal x-ray diffraction measurements on ~15 μm diameter crystals were carried at the beamline 15-ID-B of the Advanced Photon Source (APS) at Argonne National Laboratory using a wavelength of 0.41328 Å. Refinement of the single crystal data was conducted using the program SHELXL [1] after the reflections were corrected for absorption (see Ref. 2). For the space group I4/mcm, we obtained significantly worse fitting agreement ($R_1$= 8.49 %, $wR_2$= 28.0 % and Goodness of Fit =1.26). Furthermore, for the I4/mcm space group, the C-N bond distance was found to be unstable without the use of constraints.

Pair distribution function experiments were conducted at beamline the XPD (28 ID) beamline at Brookhaven National Laboratory's National Synchrotron Light Source II using a wavelength $\lambda$ = 0.18372 Å (67.486 keV). The data were measured using a Perkin Elmer detector with a sample to detector distance of 204.08 mm. The range $Q_{mim}$ = 0.23 $Å^{-1}$ and $Q_{max}$ = 25.1 $Å^{-1}$ was used in data reduction. The methods utilized for analysis of the PDF data are described in detail in Refs. [1,3]. For the fits in R-space covered the range: 2.75 < r < 60 Å. The time interval between temperature points was ~15 minutes making the experimental conditions consistent with those of the heat capacity measurements.

A single crystal was used for the x-ray absorption measurements. XAFS spectra were collected at APS beamline 13-ID-E. Four Pb L3-Edge XAFS data sets (12785 to 13788 eV) were collected in fluorescence mode using a four Vortex detectors with data corrected for deadtime. The averaged data sets were used in the analysis. Reduction of the x-ray absorption fine-structure (XAFS) data was performed using standard procedures [4]. In the XAFS refinements, to treat the atomic distribution functions on equal footing, the Pb spectra were modeled in R-space by optimizing the integral of the product of the radial distribution functions and theoretical spectra with respect to the measured spectra. Specifically, the experimental spectrum is modeled by $\chi(k) = \int \chi_{th}(r,k) 4\,\pi\, r^2 g(r) \varrho\, dr,$ where $\chi_{th}(r,k)$ is the theoretical spectrum, $\varrho$ is the bond number density. To search for asymmetry a split Gaussian distribution $4\,\pi, r^2 g(r)\varrho$ = n(r) is modeled by the function: $A\exp[-(r-r_0)^2/(2\sigma_1^2)]$ for r < $r_0$ and $A\exp[-(r-r_0)^2/(2\sigma_2^2)]$for r > $r_0$. The parameter $r_0$ is the peak position, and $\sigma_1$ and $\sigma_2$ are the left and right widths of the asymmetric Gaussian function. Theoretical spectra for atomic shells [5] were derived from the room temperature crystal structure. The k-range 2.79 < k < 11.5 $Å^{-1}$ and the R-range 1.74 < R < 3.81 Å were in the structural refinement. A value of $S_0^2$=0.9 was used (accounting for electron loss to multiple excitation channels).

Molecular dynamics (MD) simulations were conducted utilizing the VASP code implementing projector-augmented wave (PAW) potentials [6] to properly account for the bonding and atomic motion (see also Ref. [7]). The LDA exchange functional (Ceperly and Alder as parameterized by Perdew and Zunger [8]) were used with a 400 eV energy cutoff. These potentials are all standard PAW optimized potentials within VASP. A 2x2x1 orthorhombic supercell (based on the orthorhombic cell of Ref. [9]) with 192 atoms was utilized to full models the room temperature without imposing structural constraints.

For separate MD simulations, the system temperature was set at 200, 300, 400 and 500 K utilizing the (N V T) ensemble. Time steps of 0.5 fs were carried out (to properly account for the motion of atoms within the $CH_3NH_3^+$ ions). No constraints on the atomic positions were utilized. The real atomic masses (no duteration of H atoms) were used in the simulations. An equilibration run of ~3300 steps (1.5 ps) was followed by as a second set of ~3300 time-steps. The latter run was utilized in the calculating the MD-derived properties shown below. For the defect structure with an MA ion and I atom removed, the same procedure was carried out as for the standard 300 K simulations. Note that the molecular dynamics simulation are conducted assuming no specific space group (P1 space group) and imposing no specific constraints on the atomic positions.

# II. Results

**Table S1. Structural Parameters from Single Crystal Refinement at Room Temperature**

| **Atoms** | **x** | **y** | **z** | Ueq ($Å^2$)×$10^3$ |
|---|---|---|---|---|
| Pb | 0 | 0 | 0.22462(2) | 39.8(8) |
| I1 | 0 | 0 | 0.4718(13) | 121 (3) |
| I2 | 0.2780(6) | 0.2220(6) | 0.2207(17) | 112 (3) |
| N | 0.4168(42) | 0.0833(42) | 0.4520(35) | 26 (11) |
| C | 1/2 | 0 | 0.3677(78) | 76 (24) |

N site occupancy = 50%

$U_{ij}$ (Pb) 0.0390(9) 0.0413(10) 0.0000 0.0000 *
$U_{ij}$ I1) 0.1666(44) 0.0298(25) 0.0000 0.0000 *
$U_{ij}$ (I2) 0.0929(25) 0.1506(64) -0.0548(28) -0.0381(60) *

Space Group: I4cm (Z=4)
a = 8.9406 (13) Å, c = 12.6546 (25) Å, Dx = 4.071 g/$cm^3$
Measurement Temperature: 296 K
Crystal Dimensions: ~15 μm (diameter)
Wavelength: 0.41328 Å,
BASF twin parameter: 0.47(12)
Absorption Coefficient: 13.92 $mm^{-1}$
EXTI extinction parameter:0.0225(72)
F(000) = 1040
Reflections Collected: 14538
2θmax: 34.2°
$-12 \leq h < 11$, $-11 \leq k < 12$, and $-16 \leq l < 18$,
Number of Unique Observed Reflections $F_o > 4\sigma(F_o)$: 457
Number of fitting parameters: 20
Amplitude of Max Peak in Final Difference map: 2.63 e-/ $Å^3$ (N)
$R_1$ = 7.14 %, $wR_2$ = 19.0%, Goodness of Fit = 1.09

$$R_1 = \sum ||F_0| - |F_c|| / \sum |F_0|$$

$$wR_2 = \sum w(F_o^2 - F_c^2)^2) / \sum w(F_o^2)^2$$

* Atomic displacement parameters $U_{ij}$ ($Å^2$) are in the order: $U_{11}$, $U_{33}$, $U_{12}$, and $U_{23}$.

**Table S2. Bond Distances**

| **Bond Type** | **Bond Distance** (Å) |
|---|---|
| Pb-I1 | 3.127 (16) |
| | 3.200 (16) |
| Pb-I2 | 3.181 (1) |
| <Pb-I> | 3.169 |
| C-N | 1.5 (1) |

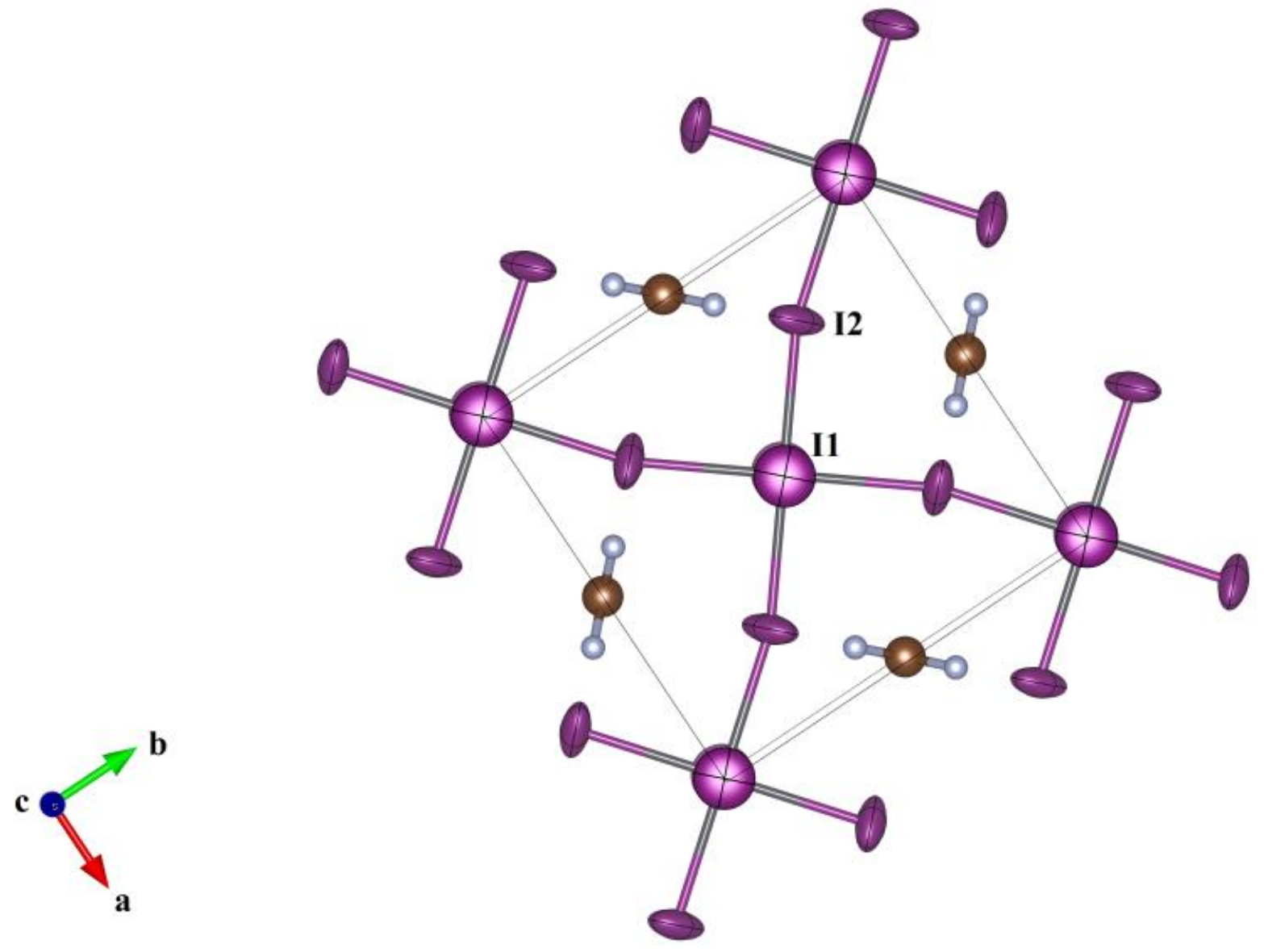

**Fig. S1.** View of the structure with the a-b plane in the plane of the figure.

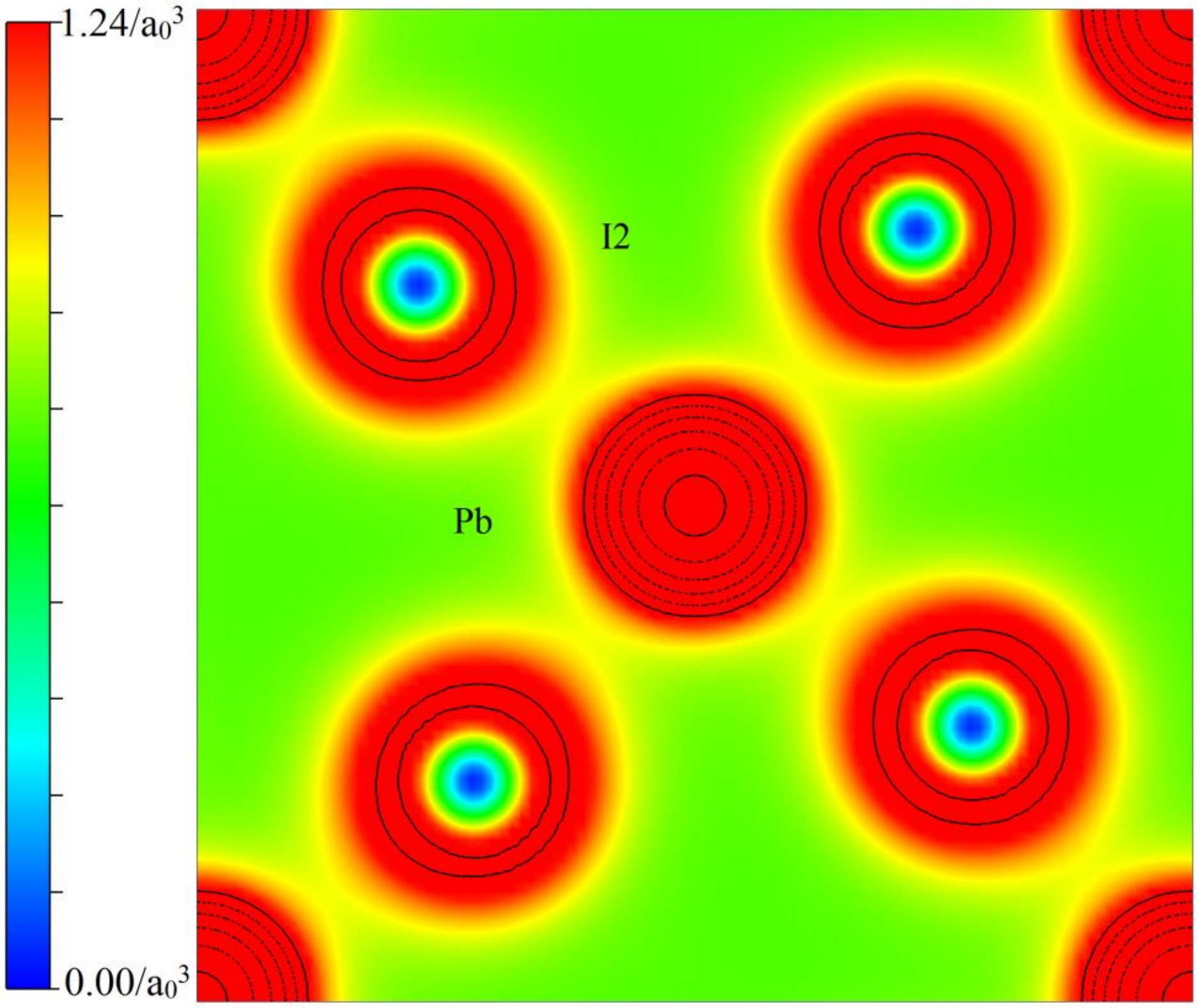

**Fig. S2.** Charged density (electrons/$a_0^3$) at z = 0.2246 based on DFT simulations using the experimental lattice parameters and atomic positions (Table S1). Note the weak charge build up between the Pb and I2 sites and the more extended distribution of the charge about the I sites compared to the Pb sites. This is distinctly different from the more localized behavior of charge on the O atoms in complex oxides (see for example Fig. 8 in Ref. [10]).

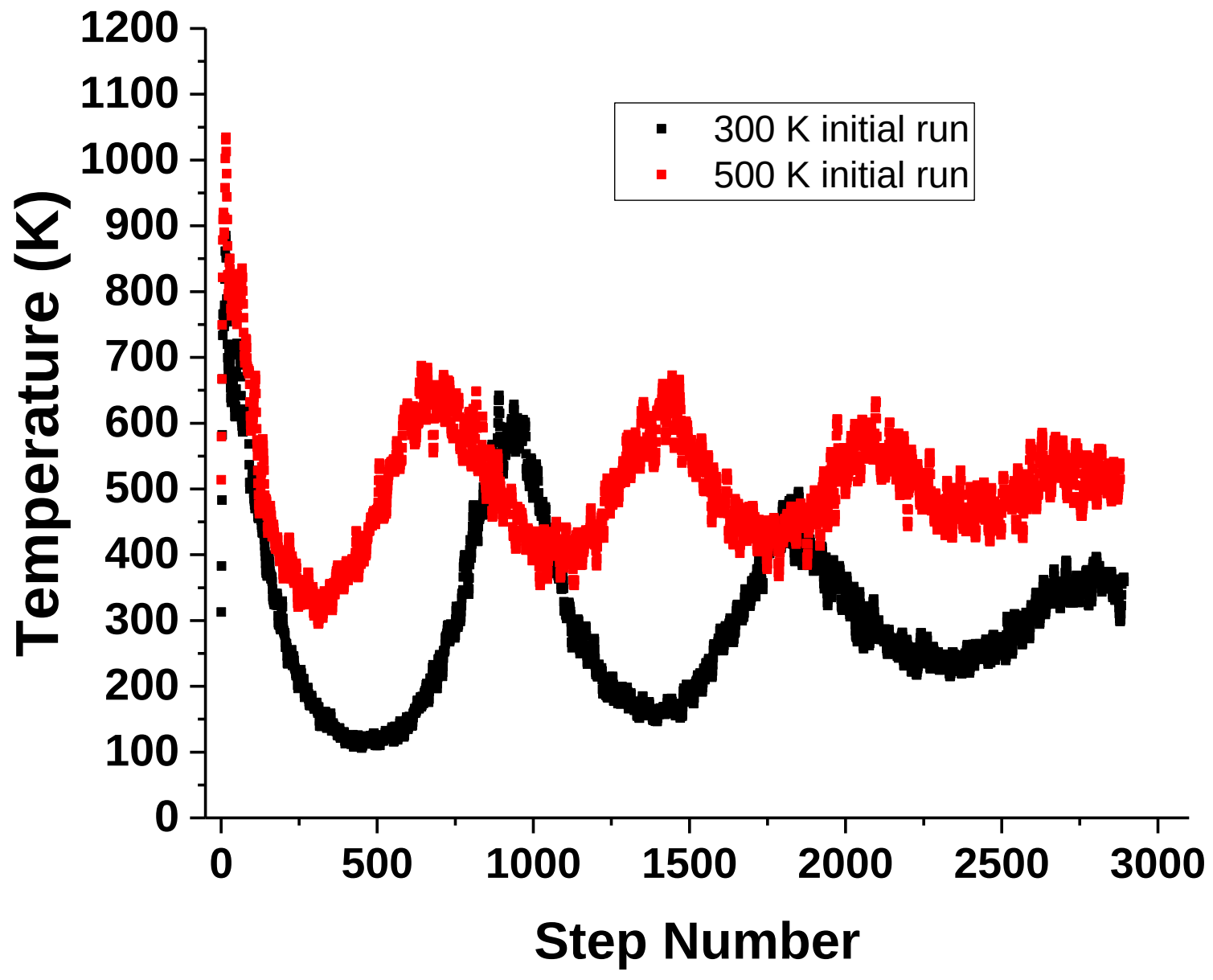

(a)

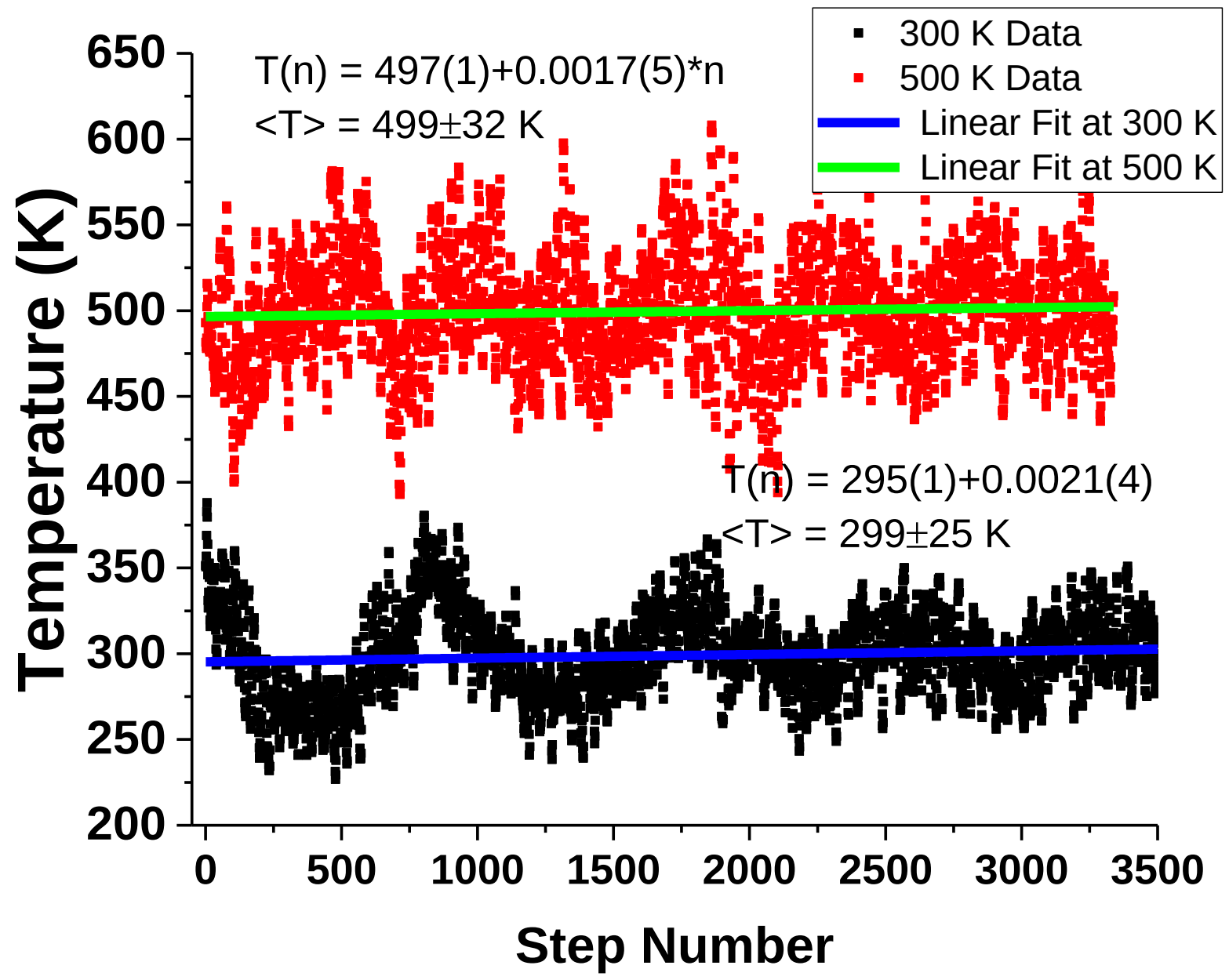

(b)

**Fig. S3.** Representative molecular dynamics simulation at 300 K and 500 K. Panel (a) gives the results from the first set of ~3000 time-steps (each step is 0.5 fs) each for the 300 K and 500 K runs. The results from the second set of ~3500 time-steps are given in (b). Thes results from these latter runs were used to compute all properties given in this work. Fits to the data as a straight line are also given to probe the level of the temperature convergence.

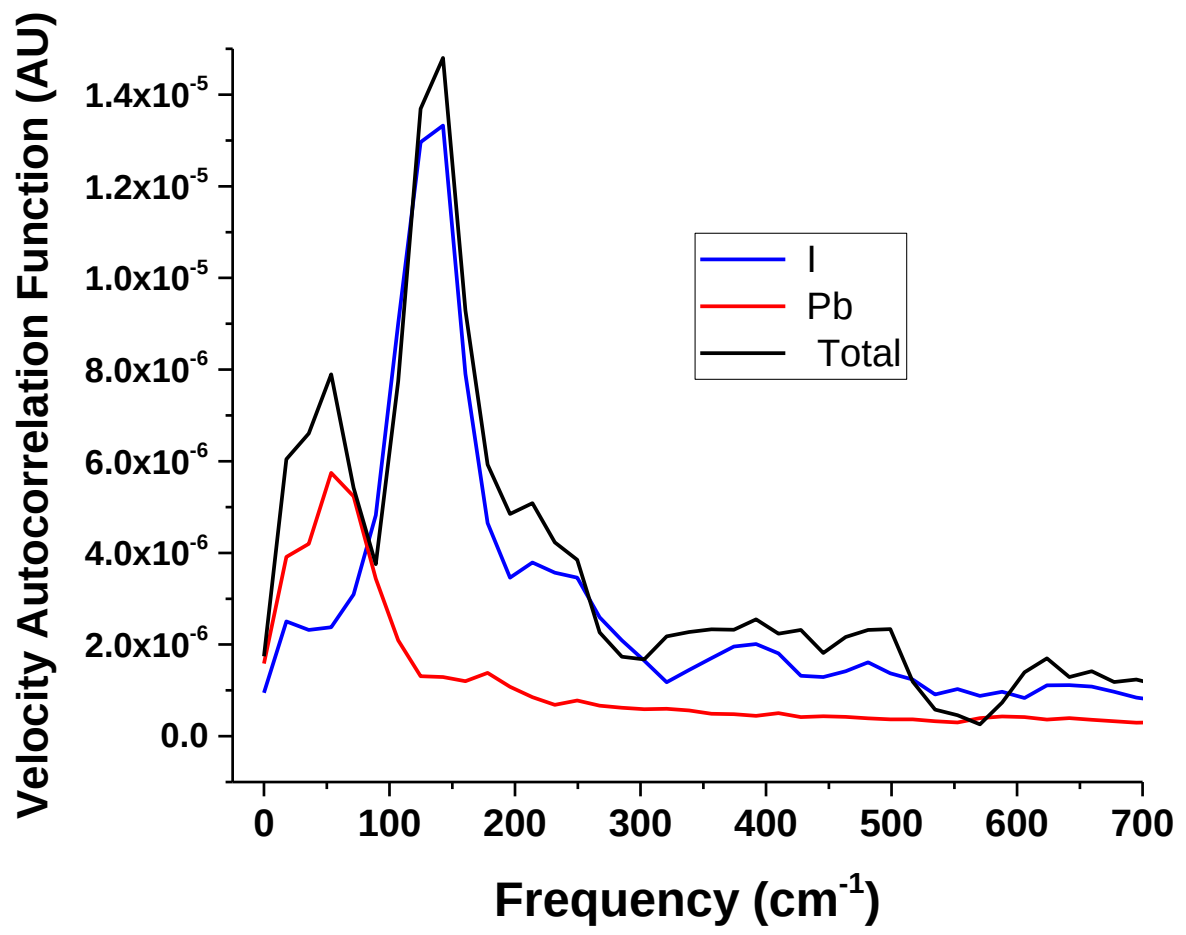

**Fig. S4.** Velocity autocorrelation function derived from the molecular dynamics simulation giving the phonon DOS for the I and Pb sites. Note the low-frequency positions for the Pb (~40 $cm^{-1}$) and I ( ~130 $cm^{-1}$) related modes in this systems. This is consistent with a mechanically soft material as found experimentally.

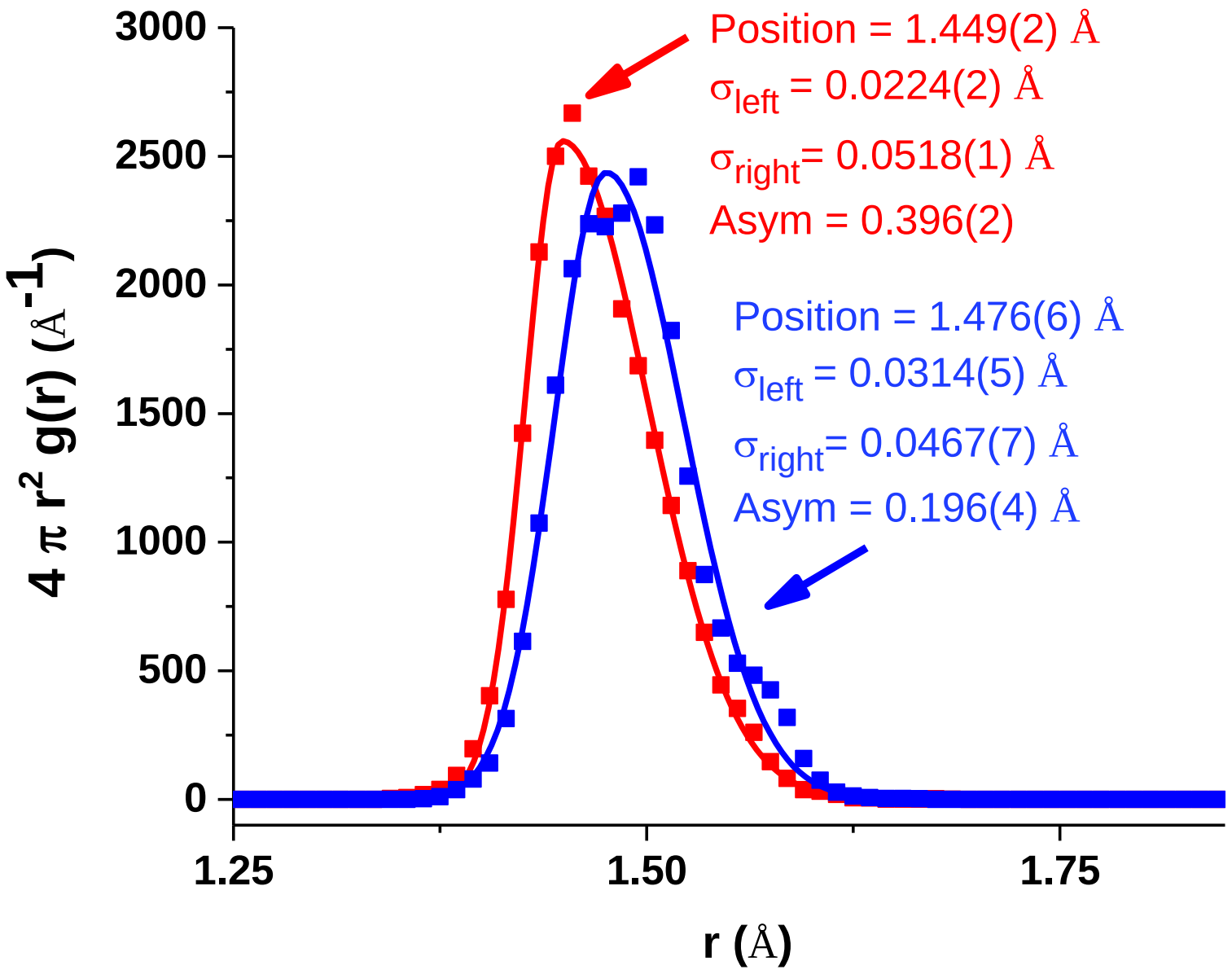

**Fig. S5.** Radial distribution function for the C-N pair of methylamine hosted in the $CH_3NH_3PbI_3$ system (red line) and in an isolated cell (same supercell without Pb and I ions and the 15,002 time-steps). Note the reduction of the asymmetry and expansion of the C-N bond length in the isolated methylamine molecule. Fits to the MD simulation with a left ($\sigma_{left}$) and right ($\sigma_{right}$) sided Gaussian (continuous at the peak) were used to assess the asymmetry (Asym $= \frac{|\sigma_{left}-\sigma_{right}|}{\sigma_{left}+\sigma_{right}}$).

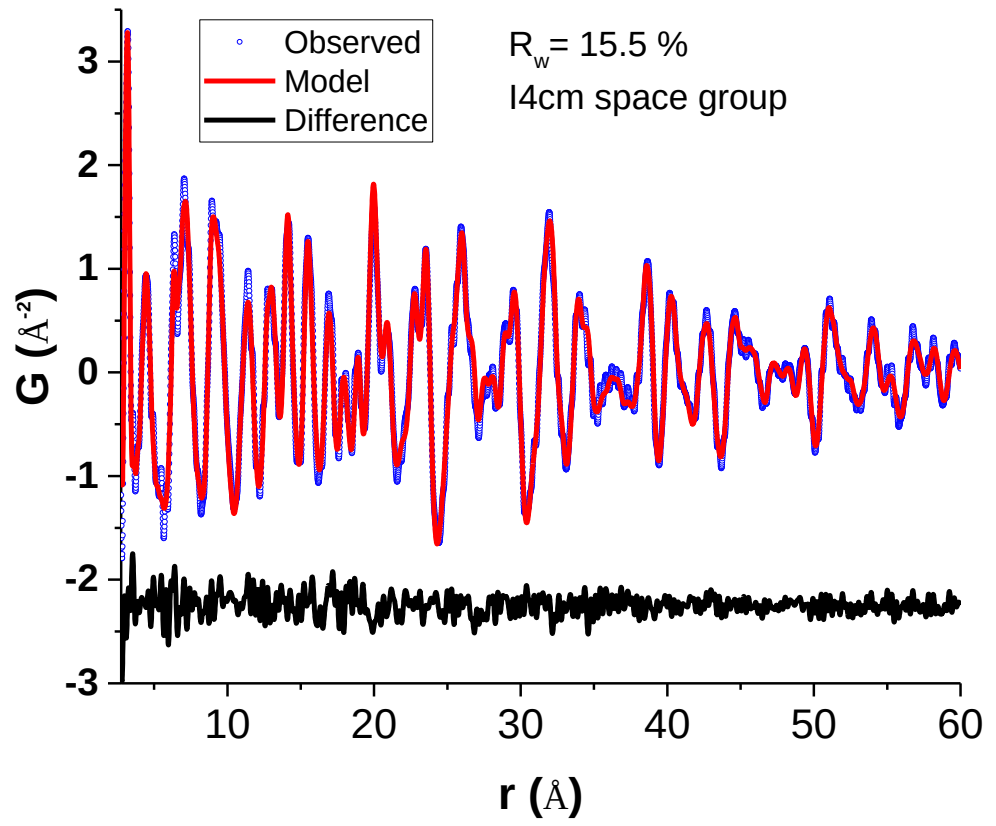

(a)

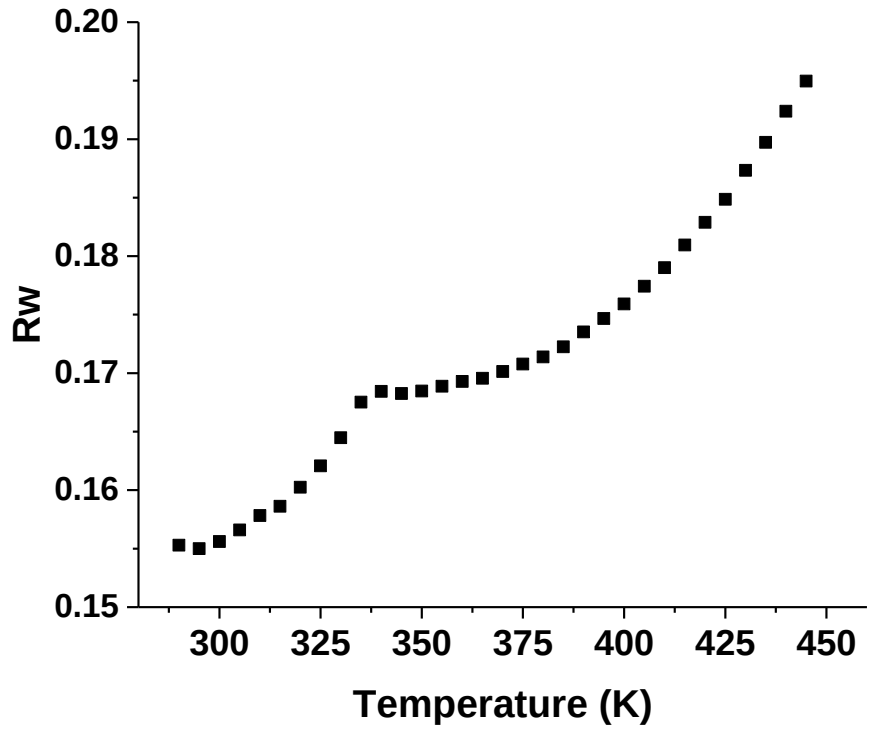

(b)

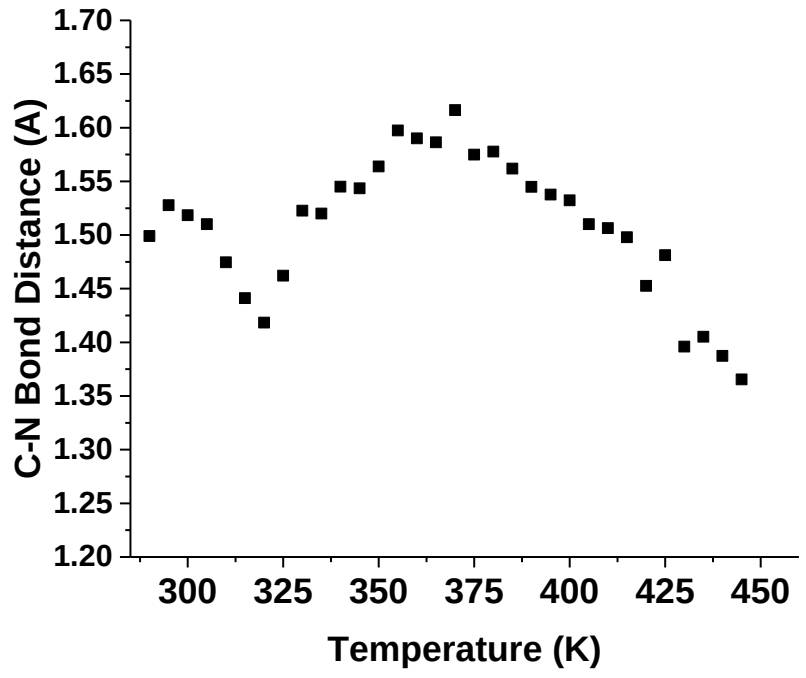

(c)

**Fig. S6.** The model derived from the room temperature single crystal structure (Table S1) was found to fit the PDF data up to 450 K. All data were fit over the range 2.75 to 60 Å to the I4cm space group model. A typical fit of the data taken at 300 K is shown in (a). $R_w$ values for the fits between 280 and 450 K are shown in (b). The extracted bond C-N bond distance is given in (c) serve to show the goodness of the fit, data quality and data calibration over the entire temperature range. The values should be compared to the results of in Fig. S3 and the results in Table S2.

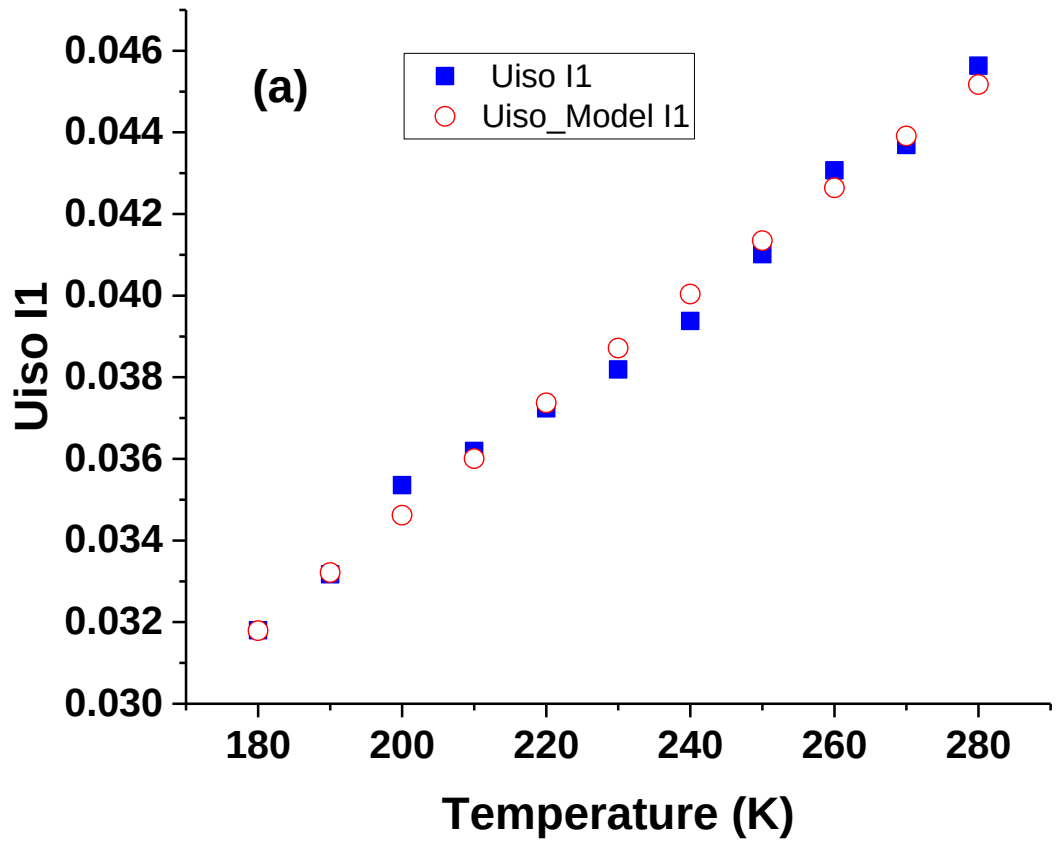

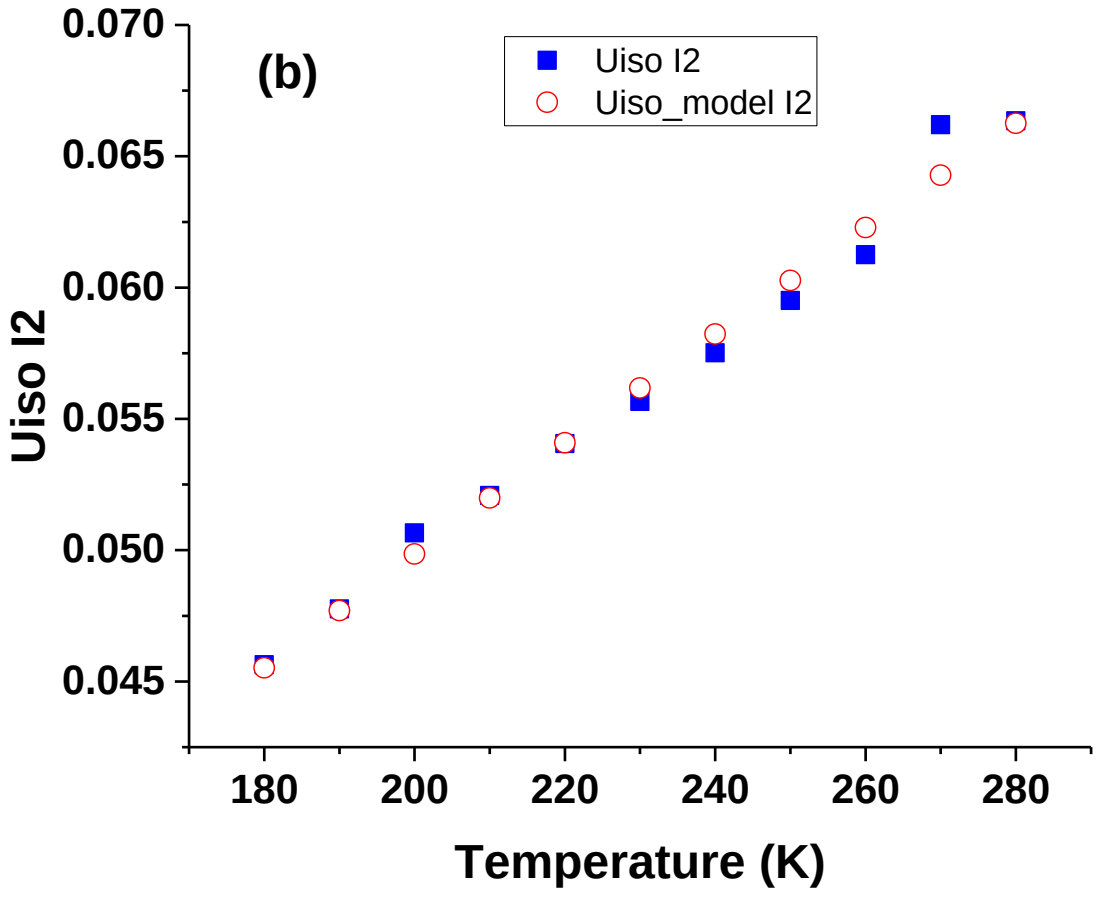

**Fig. S7.** Fits for the iodine $U_{iso}$ (Å$^2$) parameters (below T*) for I1 (a) and I2 (b) to the PDF data using a single particle potentials V(u) assuming a Boltzmann energy distribution. The potential form used was $V_0 + \alpha/2\ u^2 + \gamma\ u^4$ assuming a spherical shape of the potential. The extracted one particle potentials are given in Fig. 4(c) of the text. Closed symbols correspond to the U values from the scattering data, and open symbols are from the potential fits.

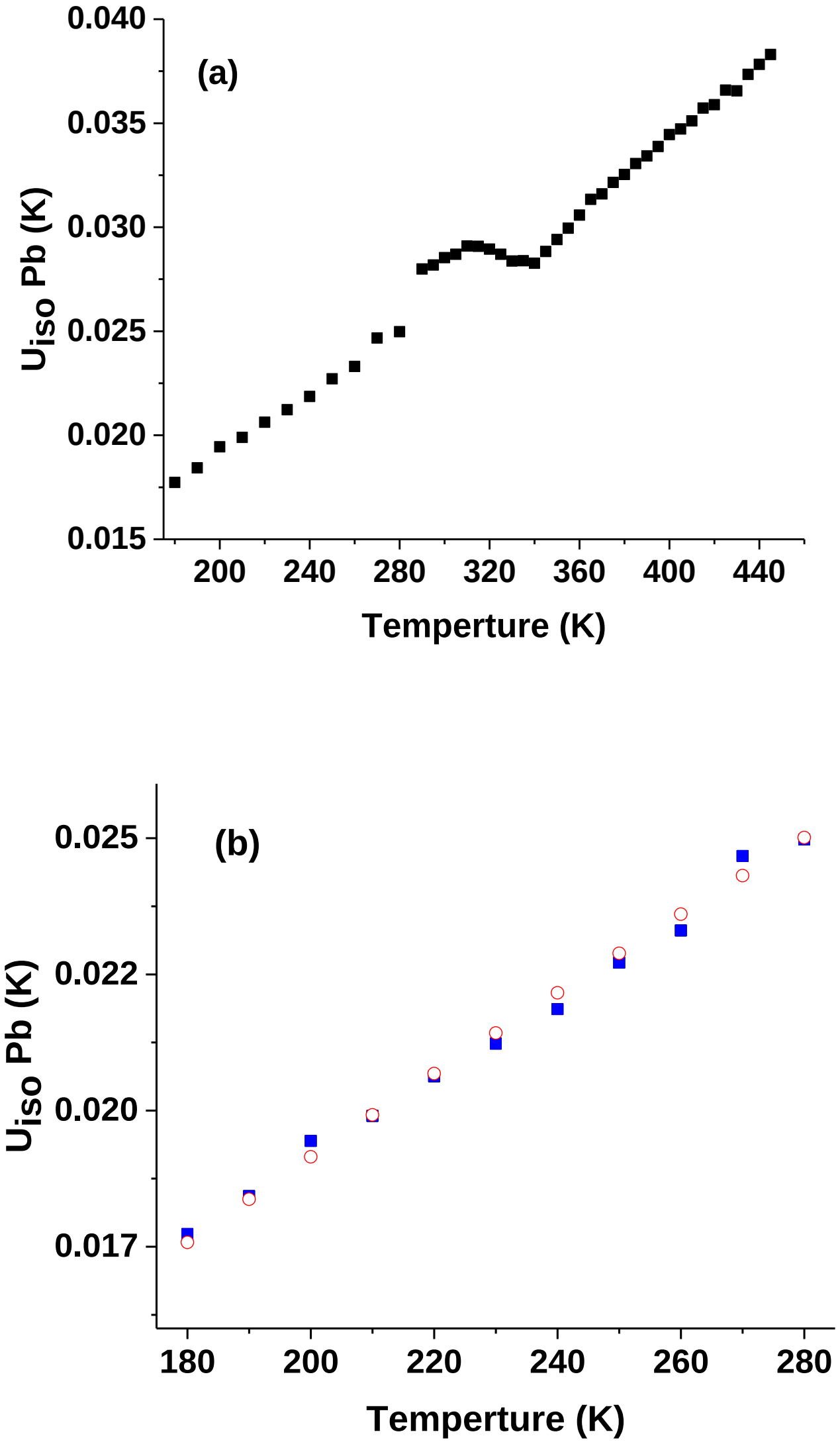

**Fig. S8.** Pb $U_{iso}$ ($Å^2$) parameters and fits below T*. (a) Thermal parameters for the Pb site showing a bump in a broad region from ~280 K to ~340 K. (b) Potential fit for Pb using the same form as that for the I sites. The symbols have the same meaning as in Fig. S7.

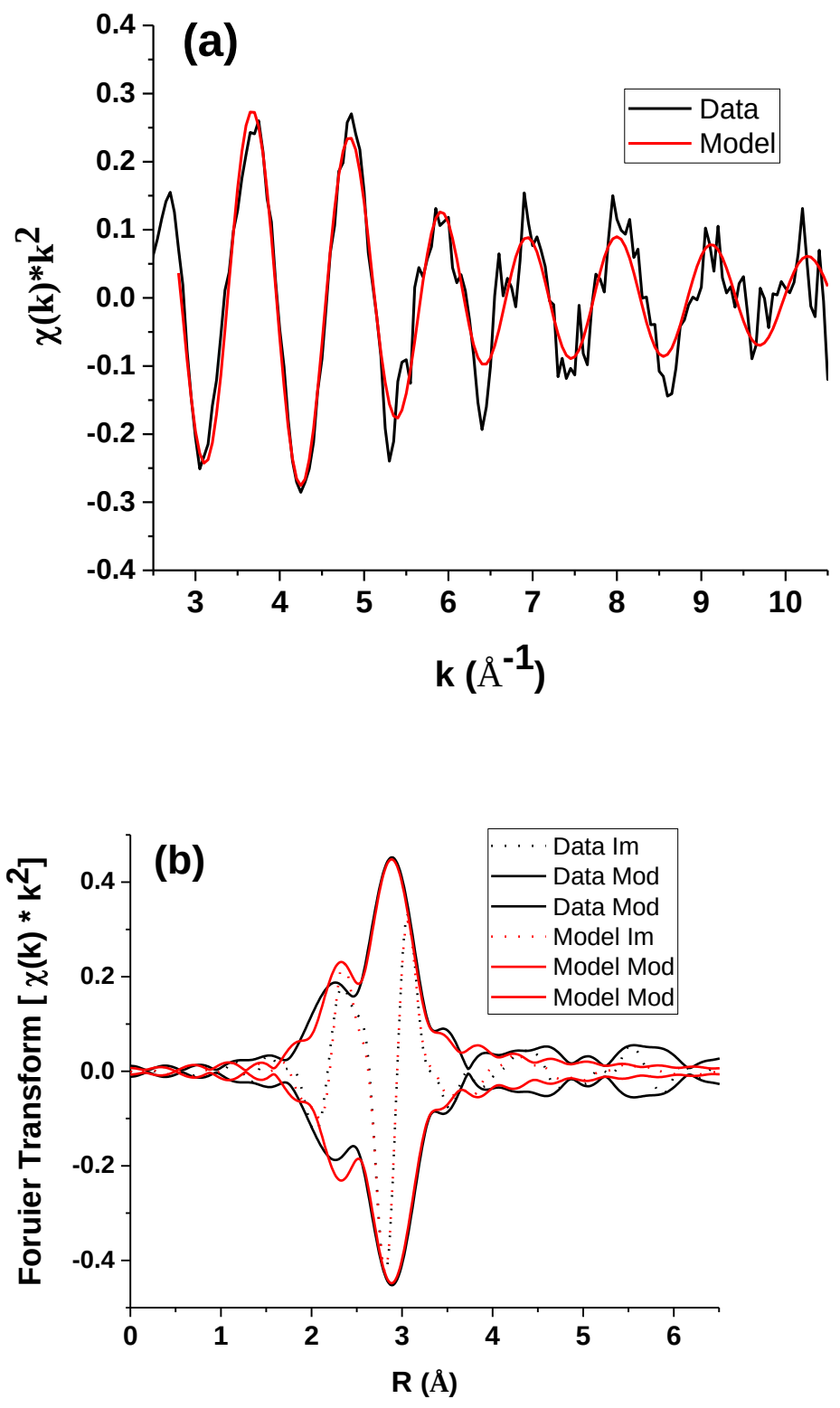

**Fig. S9.** Fits for the Pb L3 XAFS data shown in k-space (panel (a)) and in R-space (panel (b)). The fit shown is for a Pb-I bond distribution represented by an asymmetric Gaussian function as in Fig. 3(b) and Fig. S5.

# References

[1] (a) T. Yu, T. A. Tyson, P. Gao, T. Wu, X. Hong, S. Ghose, and Y.-S. Chen, Phys. Rev. B. **90**, 174106 (2014).

(b) P. Muller, R. Herbst-Irmer, A. L. Spec, T. R. Schneider, M. R. Schneider, Crystal Structure Refinement: A Crystallographer's Guide to SHELXL; Oxford Press: Oxford, 2006. (b) Sheldrick, G. M. SHELX-76: Program for Crystal Structure Determination; Cambridge University: Cambridge, 1976.

[2] SADABS: Area-Detector Absorption Correction; Siemens Industrial Automation, Inc.: Madison, WI, 1996.

[3] (a) R. B. Neder and Th. Proffen, Diffuse Scattering and Defect Structure Simulations, (Oxford University, Oxford, 2008).

(b) T. Egami and S. L. J. Billinge, Underneath the Bragg Peaks: Structural Analysis of Complex Materials, (Pergamon, Amsterdam, 2003).

(c) Th. Proffen, S. J. L. Billinge, T. Egami and D. Louca, Z. Kristallogr **218**, 132 (2003).

(d) V. Petkov, in Characterization of Materials, (John Wiley and Sons, Hoboken, 2012).

[4] (a) T. A. Tyson, M. Deleon, S. Yoong, and S. W. Cheong, Phys. Rev. B: Condensed Matter and Materials Physics **75**, 174413 (2007).

(b) B. Ravel and M. Newville, J. Synchrotron Rad. **12**, 537 (2005); *X-Ray Absorption: Principles, Applications, Techniques of EXAFS, SEXAFS and XANES*, edited by D. C. Konningsberger and R. Prins (Wiley, New York, 1988).

[5] A. L. Ankudinov and J. J. Rehr, Phys. Rev. B **56**, R1712 (1997).

[6] (a) P. E. Blöchl, Phys. Rev. B **50**, 17953 (1994).

(b) G. Kresse, and J. Joubert, Phys. Rev. B **59**, 1758 (1999).

[7] T. A. Tyson, T. Wu, H. Y. Chen, J. Bai, K. H. Ahn, K. I. Pandya, S. B. Kim and S. W. Cheong, J. Appl. Phys. **110** , 084116 (2011).

[8] J. P. Perdew and A. Zunger Rev. B **23**, 5048 (1981).

[9] M. T. Weller, O. J. Weber, P. F. Henry, A. M. Di Pumpo and T. C. Hansen, Chem. Comm. **51**. 4180 (2015).

[10] T. A. Tyson, T. Wu, K. H. Ahn, S.-B. Kim and S.-W Cheong, Phys. Rev. B: Condens. Matter Mater. Phys. **81**, 054101 (2010).